\documentclass[letterpaper, paper,11pt]{AAS}		

\usepackage{listings}
\usepackage{algpseudocode}

\usepackage{import}
\usepackage{amsthm}
\usepackage{amsmath}
\usepackage{amssymb}
\usepackage{mathptmx}
\usepackage{etoolbox}
\usepackage{url}
\usepackage{multirow}
\usepackage{multicol}

\usepackage[pdftex,breaklinks,bookmarksnumbered,linktocpage=false,hidelinks=true]{hyperref}
\usepackage{cite}
\usepackage{bm}
\usepackage{xparse}
\usepackage[section]{placeins}
\usepackage{lipsum}
\usepackage[dvipsnames]{xcolor}  

\usepackage{subfig}
\usepackage[ruled,vlined]{algorithm2e}
\usepackage{float}
\usepackage{anyfontsize}
\usepackage{bibentry}
\usepackage{pdfcomment}
\usepackage{booktabs} 

\usepackage{acronym}

\acrodef{GTA}{Graduate Teaching Assistant}
\acrodef{GRA}{Graduate Research Assistant}
\acrodef{SID}{System Identification}
\acrodef{ML}{Machine Learning}
\acrodef{KO}{Koopman Operator}
\acrodef{DMD}{Dynamic Mode Decomposition}
\acrodef{RSO}{Resident Space Object}
\acrodef{SSA}{Space Situational Awareness}
\acrodef{SDA}{Space Domain Awareness}
\acrodef{FFT}{Fast Fourier Transform}
\acrodef{ERA}{Eigen Realization Algorithm}
\acrodef{NARMAX}{Nonlinear Auto-Regressive Moving Average Model with exogenous inputs}
\acrodef{RNNs}{Recurrent Neural Networks}
\acrodef{LEO}{Low Earth Object}
\acrodef{HEO}{Highly Elliptical Object}
\acrodef{AR}{Auto-Regressive}
\acrodef{AI}{Artificial Intelligence}
\acrodef{NN}{Neural Networks}
\acrodef{SVM}{Support Vector Machine}
\acrodef{PCA}{Principal Component Analysis}
\acrodef{GMMs}{Gaussian Mixture Models}
\acrodef{FNN}{Feed-forward Neural Networks}
\acrodef{SVM}{Support Vector Machines}
\acrodef{GP}{Gaussian Processes}
\acrodef{CNN}{Convolutional Neural Networks}
\acrodef{LSTM}{Long Short-Term Memory networks}
\acrodef{OD}{Orbit Determination}
\acrodef{SGP4}{Simplified General Perturbations model}
\acrodef{SDP4}{Simplified Deep Perturbations model}
\acrodef{TLE}{Two-Line Element sets}
\acrodef{cr3bp}[CR3BP]{Circular Restricted Three Body Problem}
\acrodef{er3bp}[ER3BP]{Elliptical Restricted Three Body Problem}
\acrodef{naff}[NAFF]{Numerical Analysis of Fundamental Frequencies}
\acrodef{fma}[FMA]{Frequency Map Analysis}
\acrodef{RTS}{Rauch-Tung-Striebel}
\acrodef{EKF}{Extended Kalman Filter}
\acrodef{IOD}{Initial Orbit Determination}

\newcommand{\mbf}[1]{\mathbf{#1}}
\newcommand{\mbb}[1]{\mathbb{#1}}
\newcommand{\bsym}[1]{\boldsymbol{#1}}
\newcommand{\mcal}[1]{\mathcal{#1}}

\theoremstyle{thmstyleone}%
\theoremstyle{thmstyletwo}%

\theoremstyle{thmstylethree}%

\PaperNumber{25-577}

\begin{document}

\title{State Forecasting in an Estimation Framework with
Surrogate Sensor Modeling}

\author{Sriram Narayanan\thanks{Department of Mechanical and Aerospace Engineering,
The Ohio State University,
201 W 19th Ave,
Columbus, 43210, Ohio, USA},  
Mohamed Naveed Gul Mohamed\thanks{Department of Aerospace Engineering,
Texas A \& M University,
710 Ross St,
College Station, 77843, Texas, USA},
Ishan Paranjape\footnotemark[2], 
Indranil Nayak\thanks{SLAC National Accelerator, Stanford University, Menlo Park, 94025, CA, USA}, 
Suman Chakravorty\footnotemark[2],
\ and Mrinal Kumar\footnotemark[1]
}

\maketitle{}

\begin{abstract}
In recent years, computational power and data availability breakthroughs have revolutionized our ability to analyze complex physical systems through the inverse problem approach. Data-driven techniques like system identification and machine learning play an important role in this field, allowing us to gain insights into previously inaccessible phenomena. However, a major hurdle remains: How can meaningful information from partial measurements be extracted?
In the aerospace domain, the challenge of state estimation is particularly pronounced due to the limited availability of observational data and the constraints imposed by sensor capabilities for tracking resident space objects (RSOs). To address these limitations, advanced compensation methodologies are required. Currently, range and bearing measurements obtained from radar and optical systems constitute the primary observational tools in the space situational awareness (SSA) community. In this work, we propose a novel framework that integrates a simplified reference dynamics model with a data-driven surrogate measurement model. This fusion process leverages the strengths of both models to estimate complex dynamical behaviors under conditions of partial observability. Extensive numerical experiments were conducted across multiple datasets to validate the proposed framework. The results demonstrate its efficacy in accurately reconstructing system dynamics from incomplete measurement data. Furthermore, to ensure the robustness of the framework, an initial consistency analysis of the surrogate modeling approach is presented. By addressing the current challenges and refining the integration of data-driven techniques with traditional physics-based modeling, this framework aims to advance state estimation methodologies in the aerospace sector.
\end{abstract}

\section{Introduction}
Accurate and efficient orbit prediction is increasingly crucial for improved \ac{SSA}~\cite{blasch2017big}. The ability to precisely forecast the trajectory of \ac{RSO} is essential for conjunction analysis and collision avoidance. Traditional orbit prediction methods rely primarily on physics-based models, often lacking robustness and effectiveness in a dynamic space environment. These challenges arise due to limited information on space conditions, RSO characteristics, and the intent of maneuvering objects.  
Recent advancements in \ac{ML} provide promising solutions to address these limitations and have been explored in aerospace applications~\cite{bishop2006pattern, rasmussen2010gaussian}. For example, Girimonte and Izzo reviewed the challenges of applying artificial intelligence in space systems, including distributed intelligence, enhanced situational awareness, and decision support for spacecraft design~\cite{girimonte2007artificial}. However, a fundamental question remains: can \ac{ML} or \ac{SID} methods be used to improve forecasting reliably and consistently? There is work in the literature where Gaussian process models have been combined with conventional estimation methods such as the \ac{EKF} to generate more accurate state estimates~\cite{peng2021fusion}. In our work, we use the \ac{DMD} algorithm described in Ref.~\cite{narayanan2024predictive} to build a surrogate model and fuse the predictions in a standard \ac{EKF} framework. Since the role of the surrogate is to forecast measurements alone, there is no need to apply any physical or pseudo-physical meaning to its output to use it with the \ac{EKF}. It can simply be used with the measurement function. Orbit information fusion in \ac{SSA} has particular challenges compared to other fields: the orbit prediction duration is usually much longer than, for instance, a robotic control problem and can vary dramatically depending on the availability of observation stations. Therefore, orbit forecasting is carried out by carefully fusing the surrogate predictions with estimates from a placeholder dynamical system.
Building upon the methodologies discussed in Refs.~\cite{narayanan2022application}~\cite{narayanan2023iterative}, this work introduces a framework for forecasting with partial observations. The previously proposed iterative scheme employed a two-step process where the surrogate model and the updated state estimates were inferred from measurement data. In this work, we refine this approach by decoupling the process: the data-driven model is used solely for learning the measurements, and we introduce a reference dynamics model as a placeholder to describe the system's approximate behavior. The reference model used is typically a simpler representation and does not account for the perturbative forces that \ac{RSO}s experience on orbit. 
We examine a few different scenarios and provide insight into when the framework works best and when it does not.
In this improved framework, the surrogate model extrapolates measurements for instances when direct observations are unavailable. Concurrently, the reference dynamics model offers approximate state estimates at that time instant. These two components are then integrated within a conventional fusion framework derived from estimation theory to produce a more accurate and refined state estimate. This decoupled approach not only simplifies the modeling process but also enhances the robustness and accuracy of the forecasting system.
The proposed framework leverages the strengths of both surrogate models and reference dynamics models to address the challenges posed by partial observations. The surrogate model, trained on available measurement data, is adept at predicting measurements for time instances where direct observations are missing. This capability is crucial in scenarios where sensor data is intermittent or incomplete. 
The fusion process combines the surrogate model's measurement predictions with the reference dynamics model's state estimates using techniques from estimation theory, such as the \ac{EKF}. This approach not only enhances the accuracy of the state estimates but also improves the robustness of the forecasting system against uncertainties and noise in the measurement data. While this work focuses on the \ac{EKF}, the framework is designed to be adaptable to other filtering algorithms, ensuring flexibility and applicability across various scenarios.
The work is structured as follows: Sec. \ref{sec:proposed_approach} introduces the proposed approach, detailing the iterative scheme, the roles of the surrogate and reference dynamics models, and the fusion process. Sec. \ref{sec:results} presents the results of applying the proposed framework to various case studies, including the simple pendulum and various orbital mechanics scenarios, covering applications in the Cislunar regime as well. Sec. \ref{sec:consistency_analysis} provides a consistency analysis of the proposed method, and finally, Sec. \ref{sec:summary_5} summarizes the key contributions and implications of this work.

\section{Proposed Framework for Forecasting via Filtering}
\label{sec:proposed_approach}
\begin{figure}
    \centering
    \pdftooltip{\includegraphics[width=0.95\textwidth]{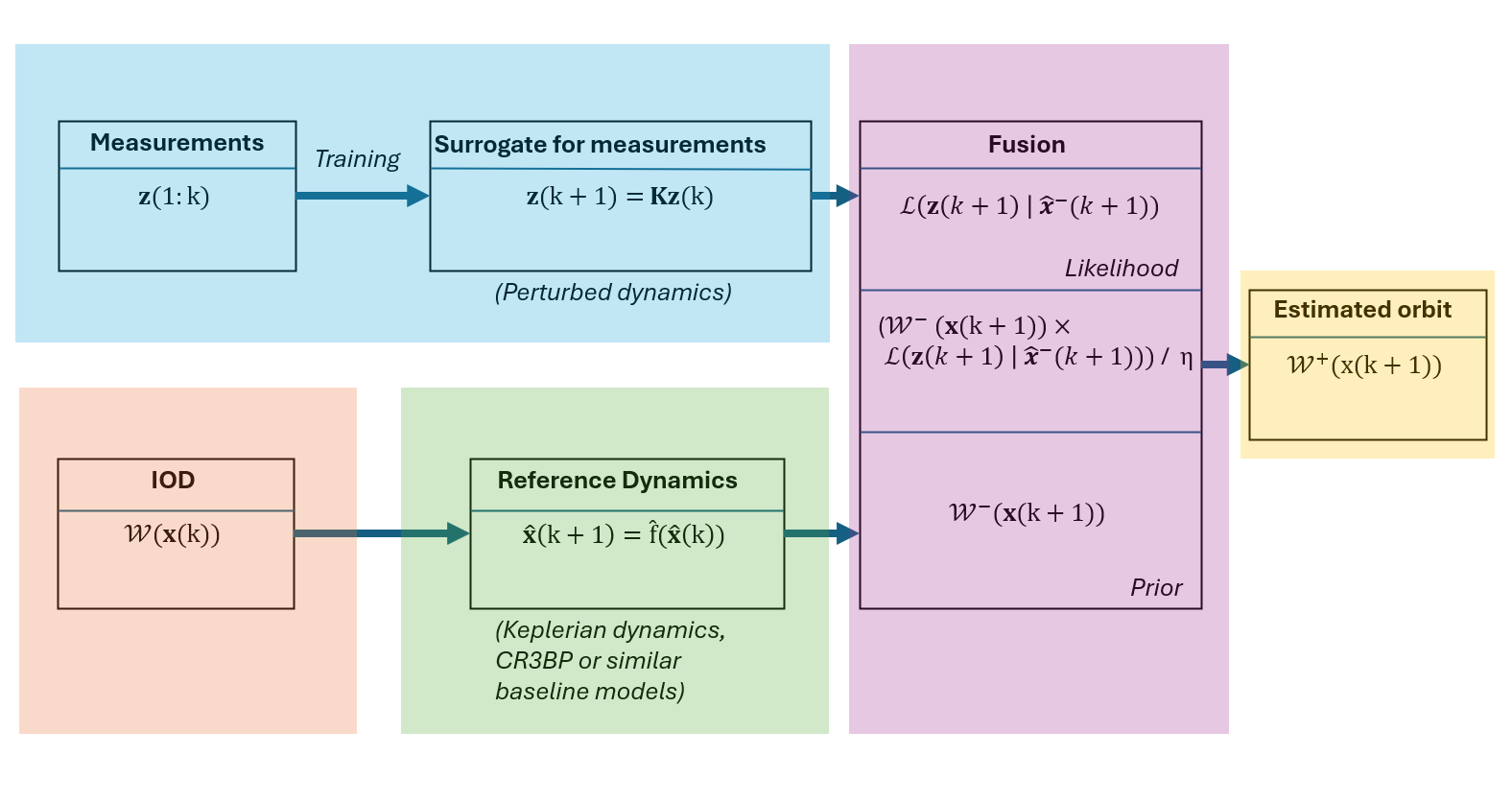}}{This figure illustrates the iterative scheme for forecasting with partial observations in an estimation framework. It includes colored zones for measurement collection, surrogate model training, initial condition determination, placeholder dynamics propagation, and state estimation fusion.}
    \caption[Iterative Scheme for Forecasting with Partial Observations in an Estimation Framework]{Iterative Scheme for Forecasting with Partial Observations in an Estimation Framework.}
    \label{fig:iter_scheme_2}
\end{figure}
In this section, we detail the proposed iterative framework for forecasting with partial observations. Fig.~\eqref{fig:iter_scheme_2} illustrates this approach. The flowchart is divided into different zones, each represented by a distinct color for clarity.
The blue zone represents the batch of measurements $\mbf{z} (1:k)$, which are accrued and used to train the surrogate data-driven model. Several different types of measurements are considered, ranging from the state vectors directly to nonlinear transformations of them, such as range, azimuth, and elevations in the context of \ac{SSA}, represented by $\mbf{z}(k) = \mbf{h}(\mbf{x}(k))$, where $\mbf{h}$ is the measurement function.
In this work, the surrogate models are constructed using the Hankel-DMD algorithm as described in Ref.~\cite{narayanan2024predictive}.
The orange zone represents the block responsible for determining the initial conditions from the collected measurements. This step is crucial for numerically integrating the reference or placeholder dynamics model, depicted in green. This process is commonly referred to as Initial Orbit Determination (IOD) in the context of orbital mechanics and space situational awareness. The batch of measurements is utilized to find a suitable initial condition, providing the initial state probability density function (pdf) $\mbf{\mcal{W}}(\mbf{x}(k))$.
The placeholder or reference model is then used to propagate the initial states to the next timestep $(k+1)$, where the data-driven surrogate has provided a new estimate for the measurement. Both sources of information are fused within \ac{EKF} framework to obtain an updated estimate for the states of the object, as depicted by the purple zone.
Finally, the yellow zone represents the updated or posterior probability density function (pdf) of the states $\mbf{\mcal{W}}^+ (\mbf{x}(k+1))$. This iterative process continues over a predefined horizon, during which the estimates from the surrogate model remain valid and useful. It is anticipated that as predictions extend beyond this range, the surrogate model will eventually fail to produce reliable results due to growing errors in the extrapolation region. The following subsections provide a detailed explanation of each representative zone.

\subsection{Hankel - DMD for learning measurements}
\label{subsec:hankel_dmd}

As discussed in Ref.~\cite{narayanan2024predictive}, according to Koopman's theory, the state-space of a nonlinear dynamical system can be transformed into an infinite-dimensional observable space where the dynamics become linear. Consider $\mbf{x}(k+1) = \mbf{f}(\mbf{x}(k))$, with a known initial condition $\mbf{x}(0)$, where $\mbf{x}(k) \in \mbb{R}^n$, and the system dynamics $\mbf{f}(\cdot)$ are unknown. The discrete-time Koopman operator (KO) $\mcal{K}$ operates on the space of observables (which could represent sensor measurements), $g(\mbf{x})$, such that the evolution of observables is described by $\mcal{K}$ as follows:
\begin{equation}
 \mcal{K} g\left(\mbf{x}(k)\right) \triangleq \left(g \circ \mbf{f}\right)\left(\mbf{x}(k)\right)=g\left(\mbf{x}(k+1)\right)
\end{equation}
The KO $\mcal{K}$ is linear but infinite-dimensional. The scalar observables $g$ are also known as Koopman observables. The eigendecomposition of the KO is given by~\cite{mohr2014construction}:
\begin{align}
    \mcal{K}\theta_m(\mbf{x}) = \lambda_m \theta_m(\mbf{x}), \ \ m = 0,1,2,....,N-1
    \label{eq:koop1}
\end{align} 
where $\theta_m(\mbf{x})$ and $\lambda_m$ are the corresponding Koopman eigenfunctions and eigenvalues. Assuming the eigenfunctions span the observable space, a vector-valued observable $\mbf{h}$ is defined in terms of the eigenfunctions:
\begin{align}
    \mbf{h}(\mbf{x}) =  \sum_{m=0}^{\infty} \theta_m (\mbf{x})\bsym{\psi}_m, \ \  \bsym{\psi}_m \in \mathbb{C}^n
    \label{eq:koop2}
\end{align}
Substituting Eq.~\eqref{eq:koop1} into Eq.~\eqref{eq:koop2}~\cite{nayak2021detection}, we get:
\begin{align}
    \mbf{h}(\mbf{f}^k(\mbf{x}(0))) = 
    \mbf{h}(\mbf{x}(k)) =  \sum_{m=0}^{\infty} \lambda_m^k \theta_m (\mbf{x}(0))\bsym{\psi}_m, \ \  \bsym{\psi}_m \in \mathbb{C}^n
\end{align}
where $\bsym{\psi}_m$ are the Koopman modes. The time expansion of the vector-valued observable $\mbf{h}$ completely determines the evolution of the nonlinear system from its initial condition. These developments allow us to capture the evolutionary mechanics of observables, which could be the partially measured state of the dynamics or other functional forms of the state (e.g., range, bearings, etc.). Despite its attractive linear structure, there is no explicit representation of the Koopman operator. Its behavior is inferred through its action on a finite-dimensional observable space. Data-driven techniques, like DMD, are typically employed to approximate the Koopman modes. DMD aims to find the most accurate approximation of $\mcal{K}$ within a finite-dimensional space. In this work, we leverage time delay embedding or Hankel reconstruction to arrange observations recursively to increase the spatial dimensions of the dataset, which is crucial for representing ergodic attractors in nonlinear dynamical systems. The Hankel matrix $\mbf{H}_k$ is constructed from observations, and higher-order DMD methods are employed to capture the system's spectral features accurately, as detailed in Ref.~\cite{narayanan2024predictive}. Once the surrogate model for the measurements is built, it is used to provide forecasts of the measurements, and these values are used in combination with the output of the placeholder model to generate updated estimates of fused measurements. The following section describes the role of the reference dynamics block in this framework.

\subsection{Reference Dynamics Model}
\label{subsec:placeholder_model}
The reference dynamics block, depicted in green in Fig.~\eqref{fig:iter_scheme_2}, contains the placeholder model that is defined based on our prior knowledge of the system and the measurements being obtained. Typically, this is a simplified analytical model that represents the system's dynamics in a rudimentary manner. For instance, in the context of \ac{SSA}, if an \ac{RSO} is being observed and measurements are being accrued, the reference dynamics block might employ a basic Keplerian two-body model. This model, while not capturing the full complexity of the \ac{RSO}'s true dynamics—such as perturbations from atmospheric drag, solar radiation pressure, or gravitational influences from other celestial bodies—serves as a "sufficiently good" estimate of the state dynamics.

The proposed framework uses this placeholder model to provide initial state estimates, which are then refined through the fusion process with the surrogate model's predictions. The surrogate model, trained on historical measurement data, captures the nuanced behavior and perturbations affecting the \ac{RSO}. By integrating the surrogate model's measurement forecasts with the reference dynamics model's state estimates within an \ac{EKF} framework, we achieve a more accurate and robust state estimation.
A key advantage of this approach is that it does not require the placeholder model to be highly accurate or complex. The model remains unchanged with its initial parameters, and the focus is on enhancing the overall state estimation through the fusion process. This decoupling of the surrogate and reference models simplifies the modeling process and ensures that the system can adapt to various scenarios without extensive modifications to the underlying dynamics model.
For example, consider a scenario involving the ISS. The reference dynamics model might use a simplified orbital model that does not account for all perturbative forces. However, the surrogate model, trained on precise tracking data, can predict the ISS's position and velocity more accurately. By fusing these predictions with the reference model's estimates, the framework can provide reliable state estimates even in the presence of partial or noisy observations. This example is discussed in detail in Sec~\ref{subsec:iss}.
This methodology is not limited to orbital mechanics. It can be applied to various fields where forecasting with partial observations is critical, such as weather prediction, financial modeling, and robotics. The flexibility and adaptability of the proposed framework make it a powerful tool for improving state estimation in complex, dynamic environments.

\subsection{Fusion of Propagated State with \textit{Predicted} Measurements}
\label{subsec:ekf_fusion}

The \ac{EKF} is a widely utilized method for state estimation in nonlinear systems~\cite{crassidis2004optimal, kalman1961new}. 
Within our framework, the EKF is employed to integrate the surrogate model's measurement forecasts with the state estimates from the reference dynamics model, ensuring a consistent and updated state representation.
The core principle of the EKF relies on the assumption that the true state remains sufficiently close to the estimated state, allowing the linearized error dynamics to provide an accurate approximation. At each iteration, the EKF propagates the state estimate and associated covariance through the nonlinear dynamics model, followed by the correction step, where incoming measurements refine the state estimate using a linearized observation model. This iterative process mitigates the impact of model uncertainties and observation noise, enhancing estimation accuracy.
Despite its advantages, the EKF's performance is inherently influenced by the validity of the linearization assumption and the choice of process and measurement noise covariance matrices. In cases where the system exhibits strong non-linearities or abrupt state transitions, the linearized approximation may introduce estimation biases. When the EKF proves inadequate, alternative filtering techniques can be employed. The Unscented Kalman Filter (UKF) addresses the limitations of linearization by using a deterministic sampling approach (unscented transform) to capture higher-order moments of the state distribution, improving estimation accuracy for highly nonlinear systems~\cite{julier2004unscented}. Another choice is the Particle Filter (PF), which provides a more general solution by representing the posterior distribution with a set of weighted particles, making it well-suited for systems with non-Gaussian noise and strong nonlinearities~\cite{ristic2003beyond}. However, the PF is computationally intensive due to its reliance on resampling techniques. When properly tuned, the EKF remains a computationally efficient and robust approach for real-time state estimation in complex dynamical systems. It serves as the primary algorithm of choice in this work.
The EKF operates in two main steps: prediction and update, as detailed in the following subsections.

\subsubsection{Prediction Step}
In the prediction step, the state estimate and the error covariance matrix are propagated forward using the reference dynamics model. Let $\hat{\mbf{x}}^+(k)$ and $\mbf{P}^+(k)$ denote the posterior state estimate and error covariance matrix at time step $k$, respectively. The prediction equations are given by:

First, propagate the state estimate $\hat{\mbf{x}}^+(k)$ forward in time using the reference dynamics model $\mbf{f}$ to obtain the predicted state estimate $\hat{\mbf{x}}^-(k+1)$,
\begin{align}
    \hat{\mbf{x}}^-(k+1) &= \mbf{f}(\hat{\mbf{x}}^+(k)) + \mbf{w}(k), \quad \mbf{w}(k) \sim \mathcal{N}(\mbf{0}, \mbf{Q}(k)).
\end{align}

Next, compute the Jacobian matrix $\mbf{F}(k)$ of the reference dynamics model $\mbf{f}$, evaluated at the posterior state estimate $\hat{\mbf{x}}^+(k)$:
\begin{align}
    \mbf{F}(k) &= \frac{\partial \mbf{f}}{\partial \mbf{x}} \bigg|_{\hat{\mbf{x}}^+(k)}.
\end{align}

The predicted error covariance matrix $\mbf{P}^-(k+1)$ is then calculated using the Jacobian matrix $\mbf{F}(k)$, the posterior error covariance matrix $\mbf{P}^+(k)$, and the process noise covariance matrix $\mbf{Q}(k)$, using the following relation,
\begin{align}
    \mbf{P}^-(k+1) &= \mbf{F}(k) \mbf{P}^+(k) \mbf{F}(k)^\top + \mbf{Q}(k).
\end{align}

\subsubsection{Update Step}
In the update step, the state estimate and error covariance matrix are updated using the surrogate model's measurement predictions. Let $\mbf{z}(k+1)$ denote the measurement prediction from the surrogate model at time step $k+1$. The update equations are given by,

First, compute the predicted measurement $\mbf{z}^-(k+1)$ using the measurement function $\mbf{h}$ applied to the predicted state estimate $\hat{\mbf{x}}^-(k+1)$ if simulating measurements. Otherwise, use the collected measurements as is:
\begin{align}
    \mbf{z}^-(k+1) &= 
    \begin{cases} 
      \mbf{h}(\hat{\mbf{x}}^-(k+1)) + \mbf{v}(k+1), & \text{if simulating measurements} \\
      \mbf{z}(k+1), & \text{otherwise},
    \end{cases}
\end{align}

Next, calculate the Jacobian matrix $\mbf{H}(k+1)$ of the measurement function $\mbf{h}$, evaluated at the predicted state estimate $\hat{\mbf{x}}^-(k+1)$,
\begin{align}
    \mbf{H}(k+1) &= \frac{\partial \mbf{h}}{\partial \mbf{x}} \bigg|_{\hat{\mbf{x}}^-(k+1)}.
\end{align}

The innovation covariance $\mbf{S}(k+1)$ is then computed using the Jacobian matrix $\mbf{H}(k+1)$, the predicted error covariance matrix $\mbf{P}^-(k+1)$, and the measurement noise covariance matrix $\mbf{R}(k+1)$,
\begin{align}
    \mbf{S}(k+1) &= \mbf{H}(k+1) \mbf{P}^-(k+1) \mbf{H}(k+1)^\top + \mbf{R}(k+1).
\end{align}

The Kalman gain $\mbf{K}(k+1)$ is then calculated using the predicted error covariance matrix $\mbf{P}^-(k+1)$, the Jacobian matrix $\mbf{H}(k+1)$, and the innovation covariance $\mbf{S}(k+1)$,
\begin{align}
    \mbf{K}(k+1) &= \mbf{P}^-(k+1) \mbf{H}(k+1)^\top \mbf{S}(k+1)^{-1}.
\end{align}

Using the Kalman gain $\mbf{K}(k+1)$, update the state estimate $\hat{\mbf{x}}^+(k+1)$ by incorporating the difference between the actual measurement $\mbf{z}(k+1)$ and the predicted measurement $\mbf{z}^-(k+1)$,
\begin{align}
    \hat{\mbf{x}}^+(k+1) &= \hat{\mbf{x}}^-(k+1) + \mbf{K}(k+1) (\mbf{z}(k+1) - \mbf{z}^-(k+1)).
\end{align}

Finally, update the error covariance matrix $\mbf{P}^+(k+1)$ using the Kalman gain $\mbf{K}(k+1)$ and the Jacobian matrix $\mbf{H}(k+1)$:
\begin{align}
    \mbf{P}^+(k+1) &= (\mbf{I} - \mbf{K}(k+1) \mbf{H}(k+1)) \mbf{P}^-(k+1).
\end{align}

\begin{table}
    \centering
    \caption{Summary of EKF Equations}
    \renewcommand{\arraystretch}{2}  
    \label{tab:ekf_summary}
    \begin{tabular}{|c|c|}  
        \hline
        \textbf{Model} & $\hat{\mbf{x}}^-(k+1) = \mbf{f}(\hat{\mbf{x}}^+(k)) + \mbf{w}(k), \quad \mbf{w}(k) \sim \mathcal{N}(\mbf{0}, \mbf{Q}(k))$ \\ 
        & $\mbf{z}^-(k+1) = \mbf{h}(\hat{\mbf{x}}^-(k+1)) + \mbf{v}(k+1), \quad \mbf{v}(k+1) \sim \mathcal{N}(\mbf{0}, \mbf{R}(k+1))$ \\ \hline

        \textbf{Covariance} & $\mbf{P}^-(k+1) = \mbf{F}(k) \mbf{P}^+(k) \mbf{F}(k)^\top + \mbf{Q}(k)$ \\ 
        & $\mbf{F}(k) = \frac{\partial \mbf{f}}{\partial \mbf{x}} \bigg|_{\hat{\mbf{x}}^+(k)}$ \\  
        & $\mbf{H}(k+1) = \frac{\partial \mbf{h}}{\partial \mbf{x}} \bigg|_{\hat{\mbf{x}}^-(k+1)}$ \\ \hline
        
        \textbf{Gain} & $\mbf{S}(k+1) = \mbf{H}(k+1) \mbf{P}^-(k+1) \mbf{H}(k+1)^\top + \mbf{R}(k+1)$ \\ 
        & $\mbf{K}(k+1) = \mbf{P}^-(k+1) \mbf{H}(k+1)^\top \mbf{S}(k+1)^{-1}$ \\ \hline
        
        \textbf{Estimate} & $\hat{\mbf{x}}^+(k+1) = \hat{\mbf{x}}^-(k+1) + \mbf{K}(k+1) (\mbf{z}(k+1) - \mbf{z}^-(k+1))$ \\  
        & $\mbf{P}^+(k+1) = (\mbf{I} - \mbf{K}(k+1) \mbf{H}(k+1)) \mbf{P}^-(k+1)$ \\ \hline
    \end{tabular}
\end{table}

The EKF framework effectively integrates the surrogate model's measurement predictions with the reference dynamics model's state estimates, thereby enhancing the accuracy and robustness of state estimation. It is worth noting that this framework is not limited to the \ac{EKF}, as it is generally compatible with any filtering algorithm that may be chosen based on the specific use case. The next section presents results from applying this framework in different situations and evaluates its effectiveness.

\section{Results}
\label{sec:results}

The evaluation criterion involved assessing the accuracy through normalized 2-norm errors and  is calculated between the true state data $\mbf{x}$ and the estimates $\bsym{\hat{x}}$ as given by, 
\begin{equation}
\bsym{\epsilon}({\mbf{x}}, \bsym{\hat{x}})= \frac{1}{k - k_0} \sum_{i=k_0}^{k} \left( \frac{\|\bsym{\hat{x}_{\text{pos}}}(i)-{\mbf{x}_{\text{pos}}(i)}\|_2}{\max(\mbf{x}_{\text{pos}})} + \frac{\|\bsym{\hat{x}_{\text{vel}}}(i)-{\mbf{x}_{\text{vel}}(i)}\|_2}{\max(\mbf{x}_{\text{vel}})} \right).
\end{equation}
Here, $\mbf{x}_{\text{pos}}$ and $\mbf{x}_{\text{vel}}$ are position and velocity states, and the index $i$ is typically set to include the extrapolated region over which predictions are made. In each scenario, a comparative evaluation is conducted against three distinct possibilities: (i) the use of a placeholder model alone; (ii) the incorporation of a DMD-based EKF fusion framework; and (iii) a hypothetical scenario where measurements are available and there is no need for data-driven surrogates to extrapolate. It is noted that in scenario (i), when using only the placeholder model, we perform a Monte Carlo propagation over an ensemble of particles to calculate and propagate the mean and covariance. This approach ensures a fair comparison against a stochastic filtering method such as the EKF. Furthermore, the ensemble size was chosen to be 200 particles, which was determined through trial and error.

\subsection{Simple Pendulum}
\label{subsec:simple_pendulum}
\begin{figure}
    \centering
    \pdftooltip{\includegraphics[width=0.55\textwidth]{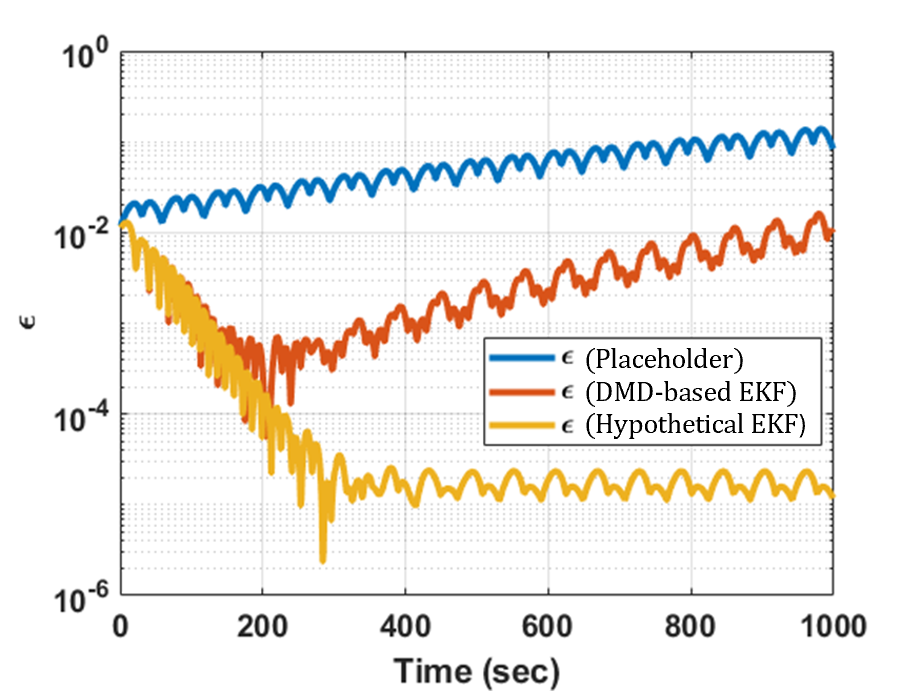}}{This figure shows the results of applying the proposed framework to a simple pendulum system with partial observations. It illustrates the effectiveness of the framework in accurately estimating the state of the pendulum.}
    \caption[Simple Pendulum Results]{Simple Pendulum Results.}
    \label{fig:spend_pobs}
\end{figure}

In this subsection, we apply the proposed framework to a simple pendulum system with partial observations. The system is modeled discretely in time, with the state vector \( \mbf{x}(k) \) at time step \( k \), where the components of \( \mbf{x}(k) \) represent the angular displacement \( \bsym{\theta}(k) \) and angular velocity \( \dot{\bsym{\theta}}(k) \). The true system parameters are a mass \( m = 0.5 \, \text{kg} \) and a damping coefficient \( c = 2 \times 10^{-3} \, \text{Ns/m} \). The placeholder system is defined using the same mass but without a damping term, resulting in a simplified reference model.

The continuous-time dynamics of the simple pendulum system are described by the second-order differential equation,
\begin{equation}
    m \ddot{\bsym{\theta}} + c \dot{\bsym{\theta}} + mgL \sin(\bsym{\theta}) = 0,
\end{equation}
where \( \bsym{\theta} \) represents the angular displacement, \( L \) is the length of the pendulum, \( g \) is the acceleration due to gravity, and \( \dot{\bsym{\theta}} \) and \( \ddot{\bsym{\theta}} \) are the first and second derivatives of \( \bsym{\theta} \) with respect to time, respectively.
For state estimation purposes, we transform this second-order differential equation into a system of first-order differential equations by defining the state vector \( \mbf{x} = \begin{bmatrix} \bsym{\theta} \\ \dot{\bsym{\theta}} \end{bmatrix} \). The continuous-time state-space representation of the system is given by,
\begin{align}
    \dot{\mbf{x}} &= \begin{bmatrix}
        \dot{\bsym{\theta}} \\
        -\frac{c}{m} \dot{\bsym{\theta}} - \frac{g}{L} \sin(\bsym{\theta})
    \end{bmatrix} + \mbf{w},
\end{align}
where \( \mbf{w} \) represents process noise. Since this is a toy example to illustrate, the process noise is set to $0$.
In our framework, we use a surrogate model to predict measurements of the pendulum's angular displacement \( \bsym{\theta}(k) \). The surrogate model is trained using the Hankel-DMD algorithm, which learns the measurements based on system dynamics observed over time. The reference dynamics model, used for comparison, is a simplified version of the pendulum's dynamics that ignores the damping term, given by,
\begin{align}
    \dot{\mbf{x}}_{\text{ref}} &= \begin{bmatrix}
        \dot{\bsym{\theta}} \\
        -\frac{g}{L} \sin(\bsym{\theta})
    \end{bmatrix}.
\end{align}
The surrogate model’s output is fused with the state estimates from the reference dynamics model using an Extended Kalman Filter (EKF). 
The results of applying this framework to the simple pendulum system are presented in Fig.~\eqref{fig:spend_pobs}. The figure demonstrates the effectiveness of the proposed approach in accurately estimating the state of the pendulum, even with partial observations and a simplified reference dynamics model. It is observed that the DMD-based EKF matches the performance of the hypothetical EKF for the first 200 seconds of the simulation but then begins to diverge due to the inability of the autoregressive modeling technique to capture dissipative effects caused by the damping force. 
Fig.~\eqref{fig:spend_pobs} also highlights that relying solely on a placeholder model for forecasting is not effective, as the error is extremely high. Therefore, this example provides evidence that using a surrogate to learn and predict measurements can serve as an intermediate solution between the hypothetical case of having constant measurements and the impractical scenario of relying solely on a placeholder model. The next section discusses the application of the proposed framework on the scenarios involving the ISS.

\subsection{International Space Station}
\label{subsec:iss}

\begin{table}[t]
\centering
\begin{tabular}{cccccc}
\toprule
$\boldsymbol{a}$ (km) & $\boldsymbol{e}$ & $\boldsymbol{i}$ (deg) & $\boldsymbol{\Omega}$ (deg) & $\boldsymbol{\omega}$ (deg) & $\boldsymbol{f}$ (deg) \\
\midrule
6796.9         & 0.0007     & 51.639         & 113.73                            & 51.197                             & 358.89        \\
\bottomrule
\end{tabular}
\caption{ISS mean orbital element set}
\label{tab:ISS_init}
\end{table}

In this section, we present applications of the proposed framework to various scenarios involving the International Space Station (ISS). The analysis begins with a baseline case featuring position-only measurements without noise, which are then used to create a surrogate measurement model. We subsequently consider several perturbation scenarios, including non-Keplerian models with J2 and drag effects, as well as Simplified General Perturbations (SGP4) models. Table~\ref{tab:ISS_init} provides the initial conditions used.
Next, we investigate a scenario with noisy observations, focusing on the non-Keplerian case with J2 perturbations. Finally, we explore an application using nonlinear observables of the states, where range, azimuth, and elevation are employed as partial observations.
In all scenarios, the placeholder dynamics model is the Keplerian model without perturbations. The results demonstrate the effectiveness of the proposed framework in accurately estimating the state of the ISS, even with partial observations and simplified reference dynamics models.

\subsubsection{Noise-free position-only observations}
\label{subsubsec:iss_position_no_noise}

\begin{figure}
    \centering
    \subfloat[Non-Keplerian with J2]{%
        \pdftooltip{\includegraphics[width=0.45\textwidth]{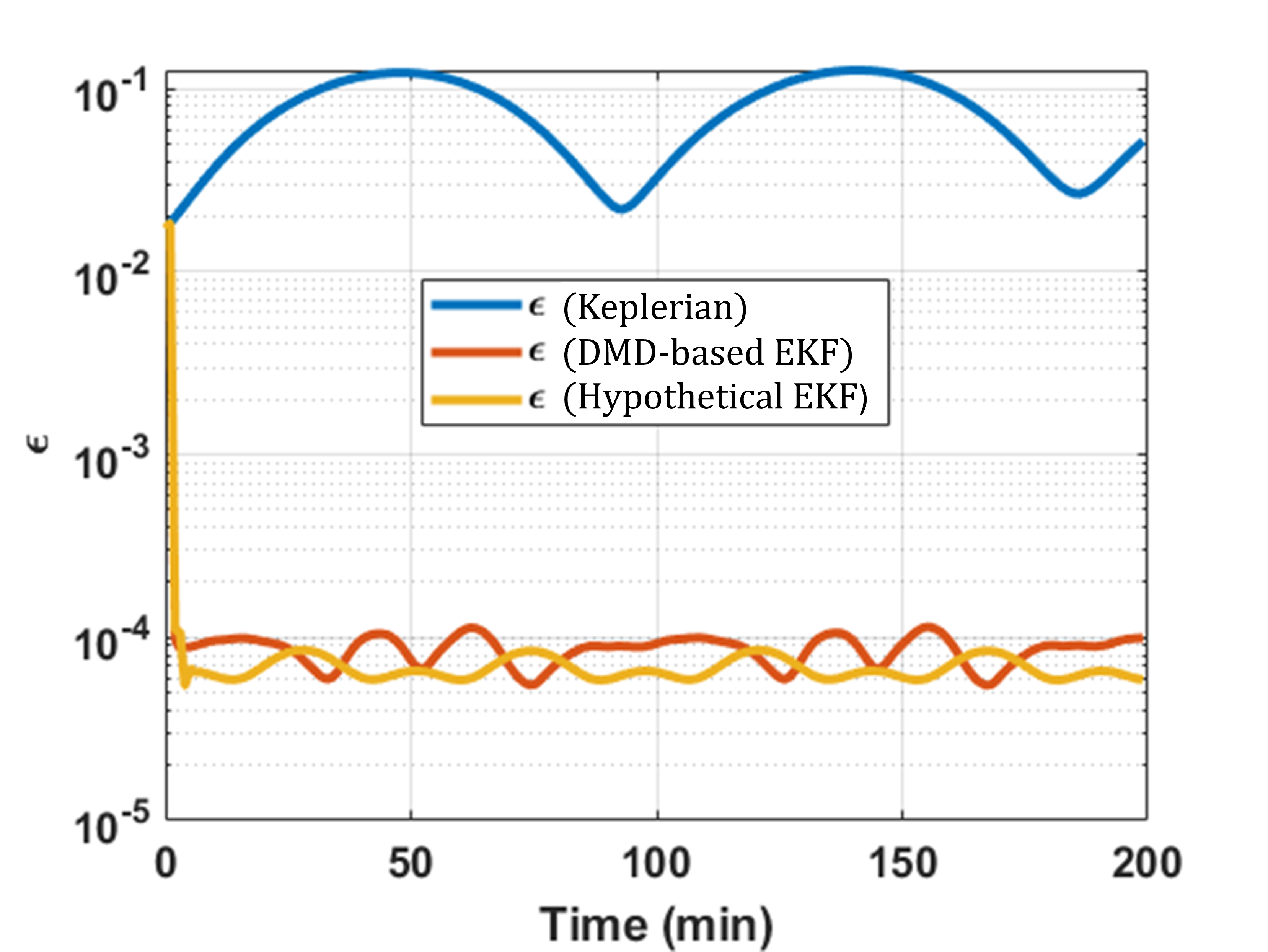}}{This figure shows the results of applying the proposed framework to the ISS with non-Keplerian dynamics including J2 perturbations.}
        \label{fig:iss_kep_j2}
    } \\
    \subfloat[Non-Keplerian with Drag]{%
        \pdftooltip{\includegraphics[width=0.45\textwidth]{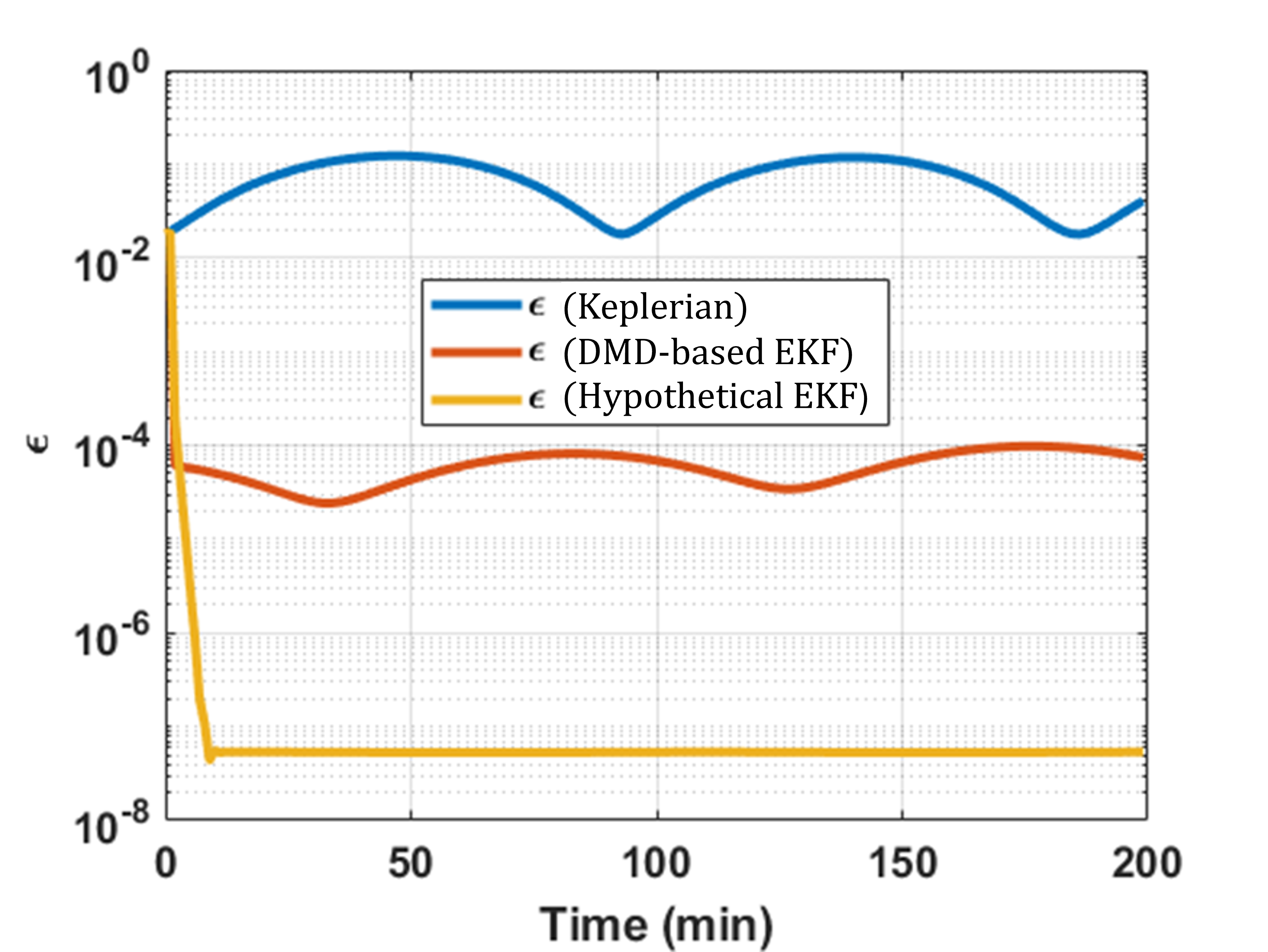}}{This figure shows the results of applying the proposed framework to the ISS with non-Keplerian dynamics including drag effects.}
        \label{fig:iss_kep_d}
    } 
    \subfloat[SGP4]{%
        \pdftooltip{\includegraphics[width=0.45\textwidth]{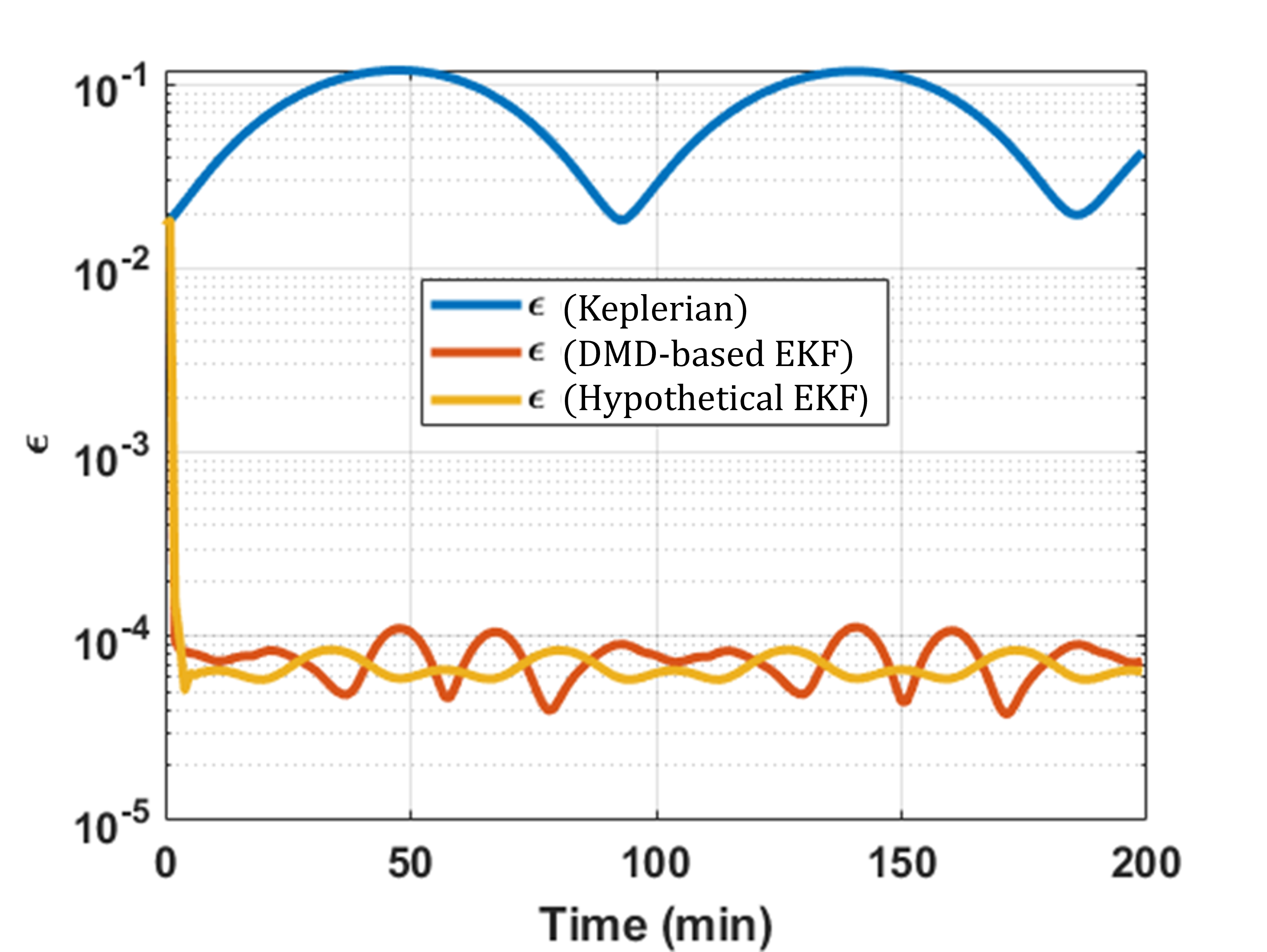}}{This figure shows the results of applying the proposed framework to the ISS using SGP4 dynamics.}
        \label{fig:iss_kep_sgp4}
    }
    \caption[ISS Position Only No Noise]{ISS Position Only No Noise. This figure compares the performance of the proposed framework under different scenarios: non-Keplerian dynamics with J2, non-Keplerian dynamics with drag, and SGP4 dynamics.}
    \label{fig:iss_position_no_noise}
\end{figure}
The same can be said for the SGP4 case, where the spectral content in the measurements is captured well, leading to the robust performance of the framework. In contrast, the case of non-Keplerian mechanics with Drag reveals that DMD has not accurately captured the spectral content. This is attributed to the method's inability to capture dissipative effects, which typically show up as a change in the rank of the reduced-order matrix $\tilde{\bsym{A}}$. Instead of one-shot learning, an improved approach would involve retraining Hankel-DMD once new measurements are received to obtain a more accurate representation of the changing spectral content.
The next section discusses a scenario similar to this one but with noisy observations. The analysis is restricted to non-Keplerian dynamics with J2 for conciseness, as the focus is on describing the impact of noise. Extensions to other scenarios can be made easily.

\subsubsection{Noisy position only observations}
\label{subsubsec:iss_position_with_noise}
\begin{figure}
    \centering
    \subfloat[Non-Keplerian with J2]{%
        \pdftooltip{\includegraphics[width=0.45\textwidth]{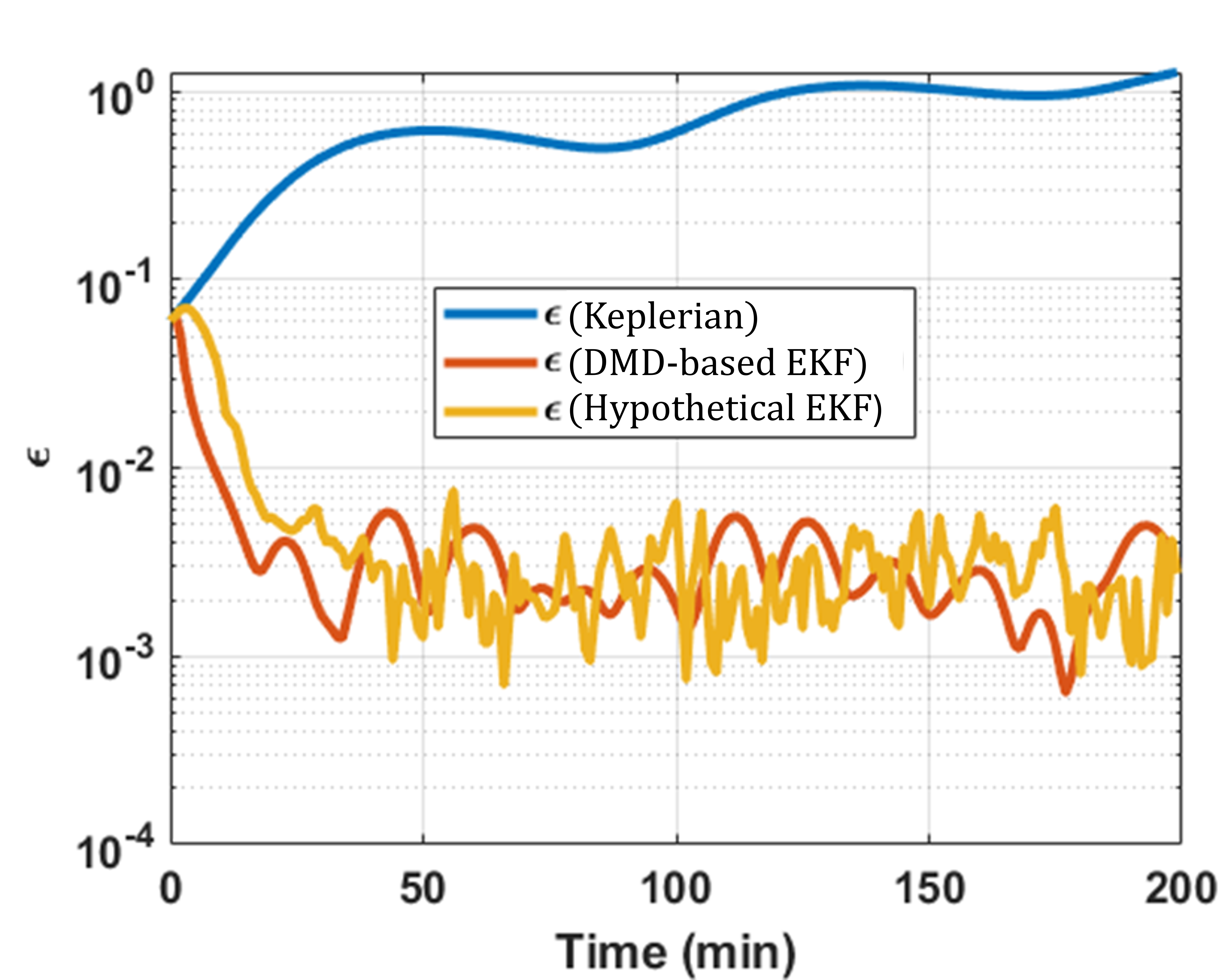}}{This figure shows the results of applying the proposed framework to the ISS with non-Keplerian dynamics including J2 perturbations and noisy position-only observations.}
        \label{fig:iss_kep_j2_pos_0}
    } \\
    \subfloat[Non-Keplerian with J2]{%
        \pdftooltip{\includegraphics[width=0.45\textwidth]{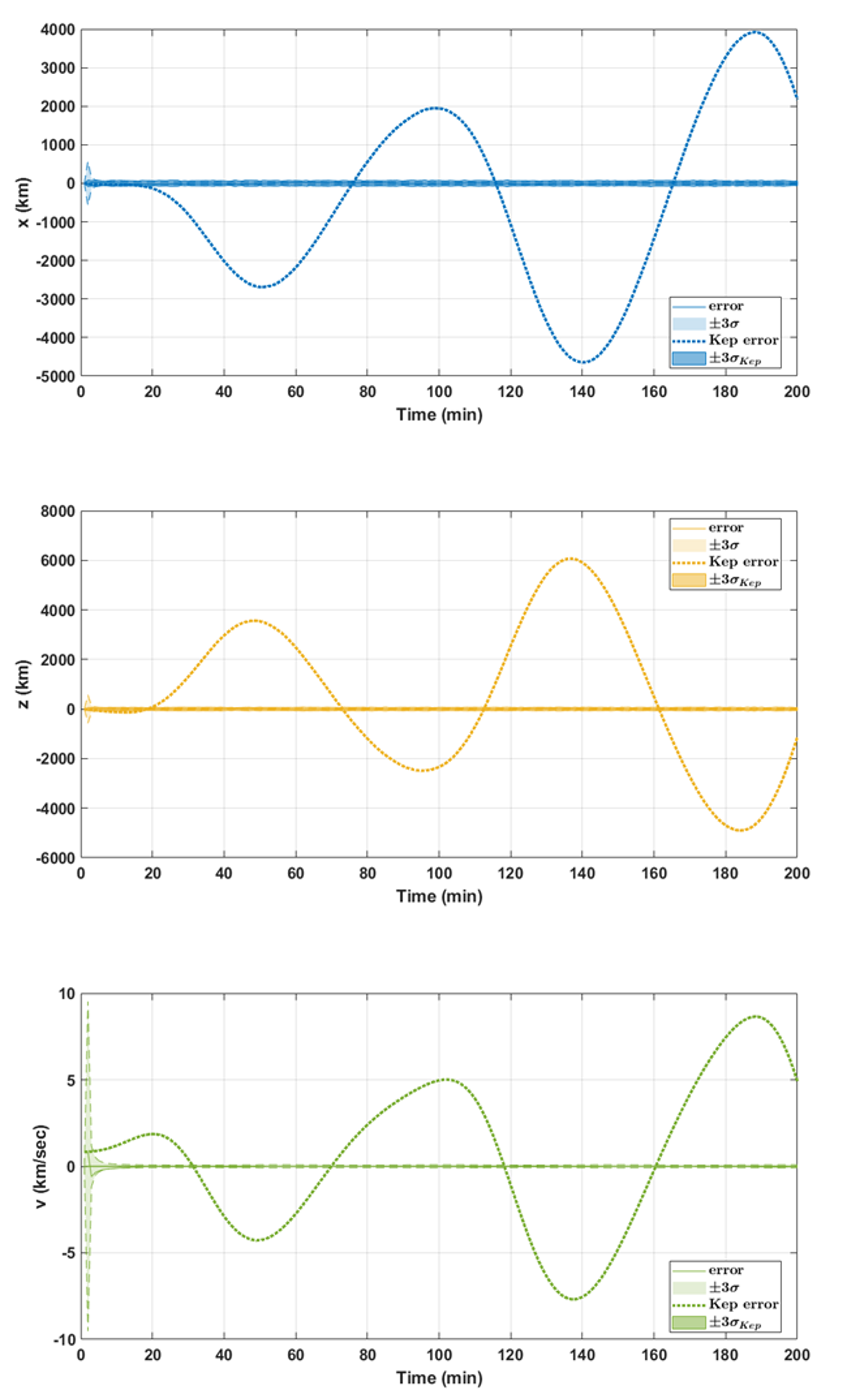}}{This figure shows the 3-sigma bounds for the x,z and v state vectors of the ISS with non-Keplerian dynamics including J2 perturbations and noisy position-only observations.}
        \label{fig:iss_kep_j2_pos_1}
    }
    \subfloat[Non-Keplerian with J2]{%
        \pdftooltip{\includegraphics[width=0.45\textwidth]{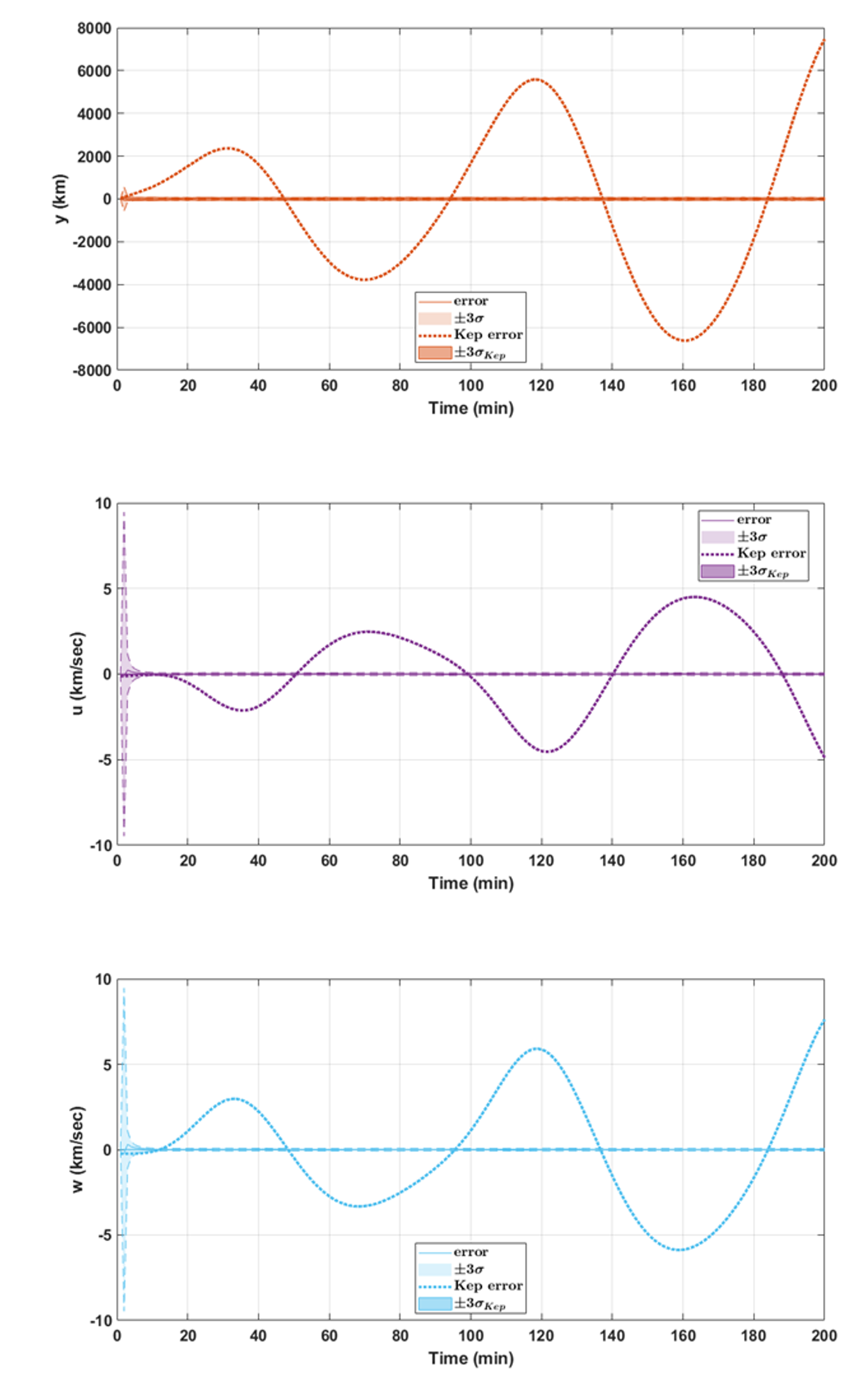}}{This figure shows the 3-sigma bounds for y, u and w state vectors of the ISS with non-Keplerian dynamics including J2 perturbations and noisy position-only observations.}
        \label{fig:iss_kep_j2_pos_n}
    }
    \caption{ISS Position Only with Noise}
    \label{fig:iss_position_with_noise}
\end{figure}
This section discusses the performance of the framework in the context of the ISS with noisy position-only observations. The true dynamics are assumed to be non-Keplerian with J2 perturbations and an added sensor noise of $5\%$, while the placeholder system remains the same Keplerian model free of perturbations. Fig.~\eqref{fig:iss_kep_j2_pos_0} depicts the results. Even in the presence of noise, albeit small, DMD can learn an accurate representation of the measurement dynamics by effectively filtering out the noisy high-frequency modes. When combined in the filtering framework with a good initial guess for the covariance and process noise matrices, the results are favorable. It is assumed that the sensor noise matrix $\mbf{R}$ is known exactly. Figs.~\eqref{fig:iss_kep_j2_pos_1}\eqref{fig:iss_kep_j2_pos_n} show the $3-\sigma$ bounds for each of the six state vectors, $\mbf{x,y,z,u,v,w}$ being the three position and three velocity states respectively. It is seen that the estimates from the framework lie within this region; however, if the Keplerian model is used directly, the estimates are widely diverging, further implying the need for this method when measurements may not be consistently available. The next section discusses a more realistic scenario where range, azimuth, and elevation are used as observations.

\subsubsection{Range, Azimuth, Elevation measurements}
\label{subsubsec:iss_rae}
\begin{figure}
    \centering
    \pdftooltip{\includegraphics[width=0.55\textwidth]{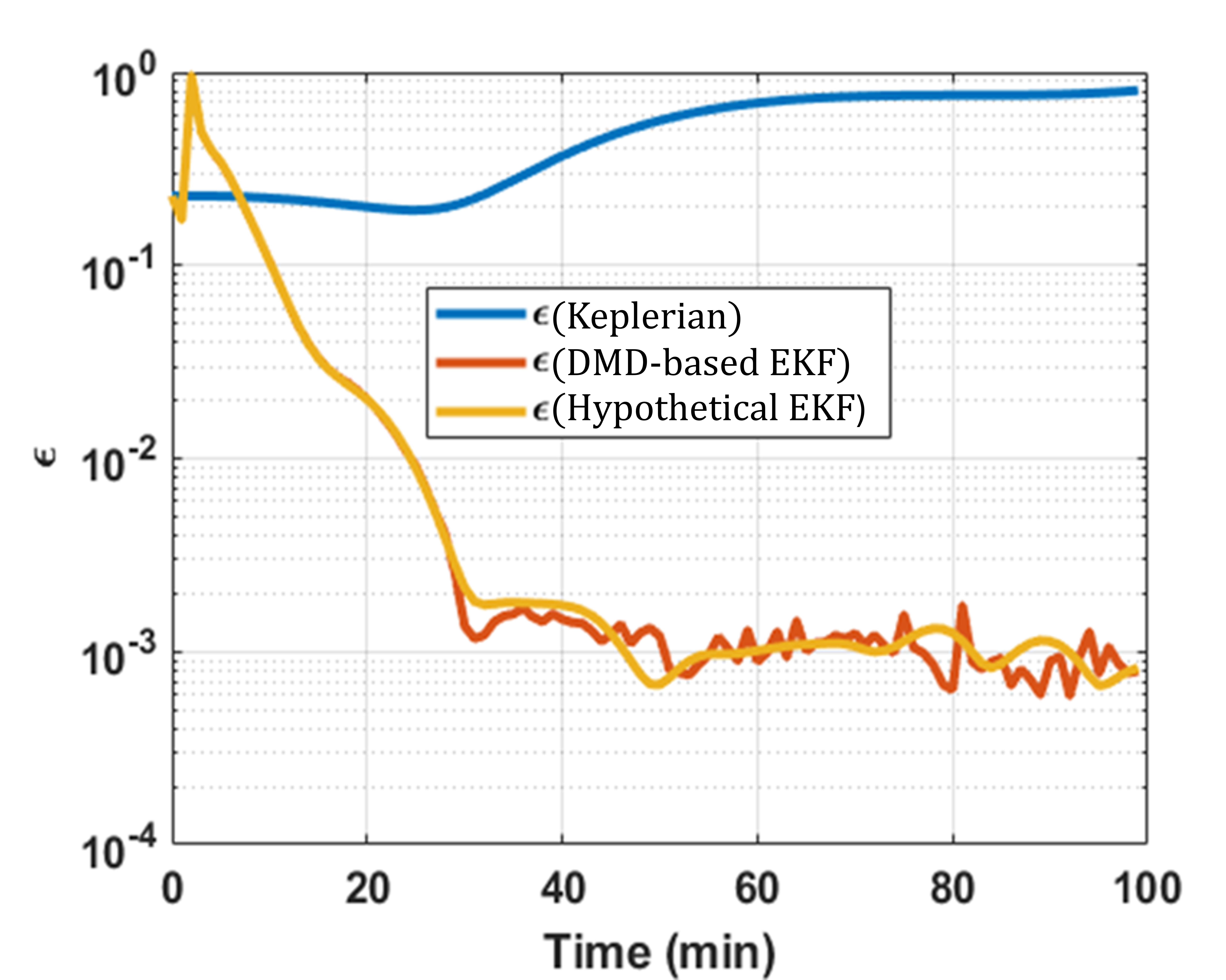}}{This figure shows the results of applying the proposed framework to the ISS with range, azimuth, and elevation measurements under non-Keplerian dynamics including J2 perturbations.}
    \caption{ISS Range, Azimuth, Elevation Measurements - Non-Keplerian with J2}
    \label{fig:iss_kep_j2_rae}
\end{figure}
This section discusses the performance of the framework when applied to nonlinear transformations of the state variables. Here, the observables considered are range, azimuth, and elevation. For simplicity, the observer is located in Columbus with the corresponding longitude, latitude, and altitude and uses a sensor with an infinite field of view, i.e., the observer can track the object at all times. The focus here is to demonstrate the framework's effectiveness in scenarios where noisy nonlinear observables are considered. A more realistic field of view may be used for the sensor; however, an appropriate interpolation scheme is required in this case, typically some sort of polynomial basis method. However, the measurements will then be filtered/averaged versions and will lack the noise characteristics. The goal is to demonstrate on noisy nonlinear observables; hence, the scenario is set up in such a manner. Fig.~\eqref{fig:iss_kep_j2_rae} depicts the results. In this scenario, the observables have a standard deviation of 1 km in range and 0.01 radians in the azimuth and elevation. The initial conditions are determined using the Herrick-Gibbs method~\cite{crassidis2004optimal}.
Here again, it can be seen that the framework matches performance with the hypothetical case. This is attributed to the nonlinear observables, in this case, of the periodic states being periodic themselves, thus allowing Hankel-DMD to accurately capture the spectral content and produce reliable results on fusion.

\subsection{Limitations and Challenges}
\label{subsec:limitations_challenges}

While the proposed framework demonstrates potential in improving state estimation with partial observations, it is not without limitations and challenges. Some of the key issues are discussed below:
\begin{enumerate}
    \item \textit{Dependence on Surrogate Model Accuracy}: The performance of the framework heavily relies on the accuracy of the surrogate model. If the surrogate model fails to capture the underlying dynamics accurately, the overall state estimation will be compromised. This is particularly evident in scenarios with strong non-linearities or dissipative effects, where the surrogate model may struggle to provide reliable predictions. This is evident from the results presented in Sec.~\ref{subsec:simple_pendulum}.
    \item \textit{Noise Sensitivity}: While the framework can handle noisy observations to some extent, the presence of significant noise can still impact the accuracy of the surrogate model and the overall state estimation. Proper tuning of the process and measurement noise covariance matrices is crucial for maintaining robustness against noise. In our work, the covariance $\mbf{P}(0)$ and process noise $\mbf{Q}$ are initialized by performing a sweep via trial and error. 
    \item \textit{Initial Condition Sensitivity}: The accuracy of the initial state estimates plays a critical role in the performance of the framework. Poor initial estimates can lead to divergence in the state estimation process, especially in highly non-linear systems. In the case where there are only position measurements, the initial conditions for velocity may be determined via a finite difference approach. In the case of nonlinear observables, as in Sec.~\ref{subsubsec:iss_rae}, an appropriate IOD scheme may be utilized to generate the initial state conditions.
    \item \textit{Limited Extrapolation Capability}: The surrogate model's ability to extrapolate beyond the training data is limited. As predictions extend further into the future, the surrogate model's accuracy may deteriorate, leading to increased errors in state estimation. This limitation is standard to any scenario and can only be overcome by introducing new measurements and retraining Hankel-DMD.
    \item \textit{Model Assumptions}: The framework assumes that the reference dynamics model, although simplified, provides a reasonable approximation of the system's behavior. In cases where the reference model is significantly different from the true dynamics, the fusion process may not yield accurate state estimates.
\end{enumerate}

To illustrate these limitations and challenges, we present two case studies: (1) the application of the proposed framework to the Molniya-3-50 satellite and (2) the application to the Van Der Pol oscillator.

\subsubsection{Molniya-3-50}
\label{subsubsec:ml350}

\begin{table}[t]
\centering
\begin{tabular}{cccccc}
\toprule
$\boldsymbol{a}$ (km) & $\boldsymbol{e}$ & $\boldsymbol{i}$ (deg) & $\boldsymbol{\Omega}$ (deg) & $\boldsymbol{\omega}$ (deg) & $\boldsymbol{f}$ (deg) \\
\midrule
26555.94 & 0.7294 & 63.324 & 295.46 & 282.69 & 357.32 \\
\bottomrule
\end{tabular}
\caption{MOLNIYA-3-50 mean orbital element set}
\label{tab:ML350_init}
\end{table}

\begin{figure}
    \centering
    \pdftooltip{\includegraphics[width=0.55\textwidth]{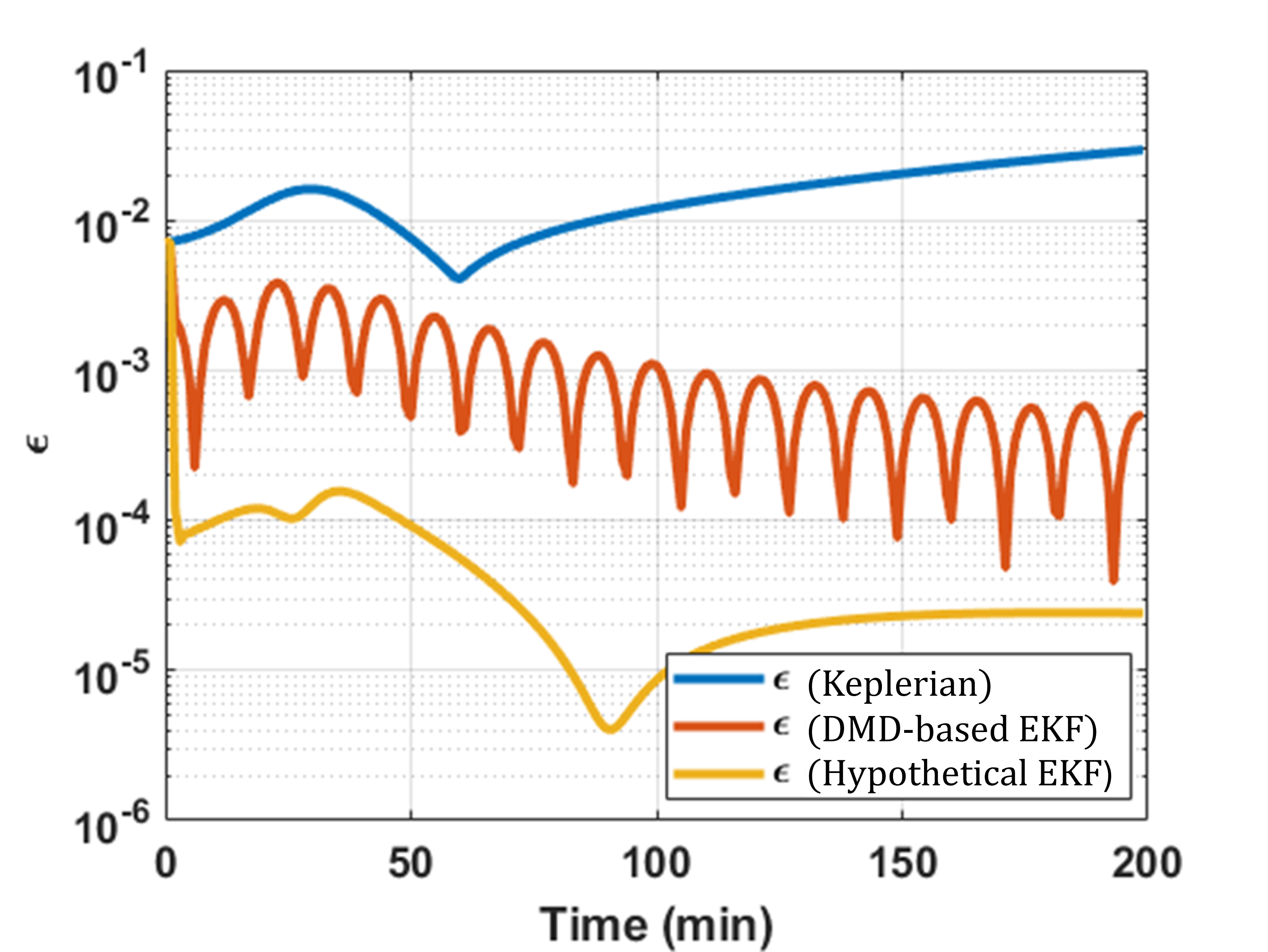}}{This figure shows the results of applying the proposed framework to the Molniya-3-50 object using the SDP4 model as the true system and a Keplerian model as the placeholder.}
    \caption{ML350 Results - SDP4}
    \label{fig:ml350_sgp4}
\end{figure}
This section highlights the issues of \textit{Dependence on Surrogate Model Accuracy} and \textit{Limited Extrapolation Capability} by considering the application of this framework on the Molniya-3-50 object. Table~\ref{tab:ML350_init} provides the initial conditions. The SDP4 model is used as the true system with noise-free position-only observations, and a Keplerian model is used as the placeholder. Similar to the above cases, the EKF is used to perform fusion and generate estimates. Here, Hankel-DMD struggles to capture the spectral content of the Molniya satellite, which is attributed to the orbit's eccentricity and the time scales involved. A more detailed discussion on this issue can be found in Ref.~\cite{narayanan2024predictive}.
In the case of the ISS, $10$ periods correspond to approximately $15$ hours, whereas for MOLNIYA-3-50, it equates to roughly $5$ days. This discrepancy allows ample time for perturbations to impact the trajectory. The perturbations cause drift in the dynamics, making it difficult for DMD to approximate the frequency spectrum accurately. The comparison between ISS and MOLNIYA-3-50 illustrates the substantial difference in applications and how the use of DMD demands a meticulous, case-specific approach rather than a generalized one. This discrepancy arises because, for a fixed window length, traversing that window through the dataset results in varying frequency content. This issue translates to the framework being unable to provide accurate estimates due to the surrogate model's inability to capture the measurements accurately, and the extrapolation remains inaccurate as well. Fig.~\eqref{fig:ml350_sgp4} depicts the same. We see that the DMD-based EKF method performs about two orders of magnitude worse than the hypothetical EKF, where the true measurements are available. The following section discusses the application on the Van Der Pol oscillator.

\subsubsection{Van Der Pol (VDP) oscillator}
\label{subsubsec:vdp_oscillator}
\begin{figure}
    \centering
    \subfloat[VDP Oscillator Results]{%
        \pdftooltip{\includegraphics[width=0.45\textwidth]{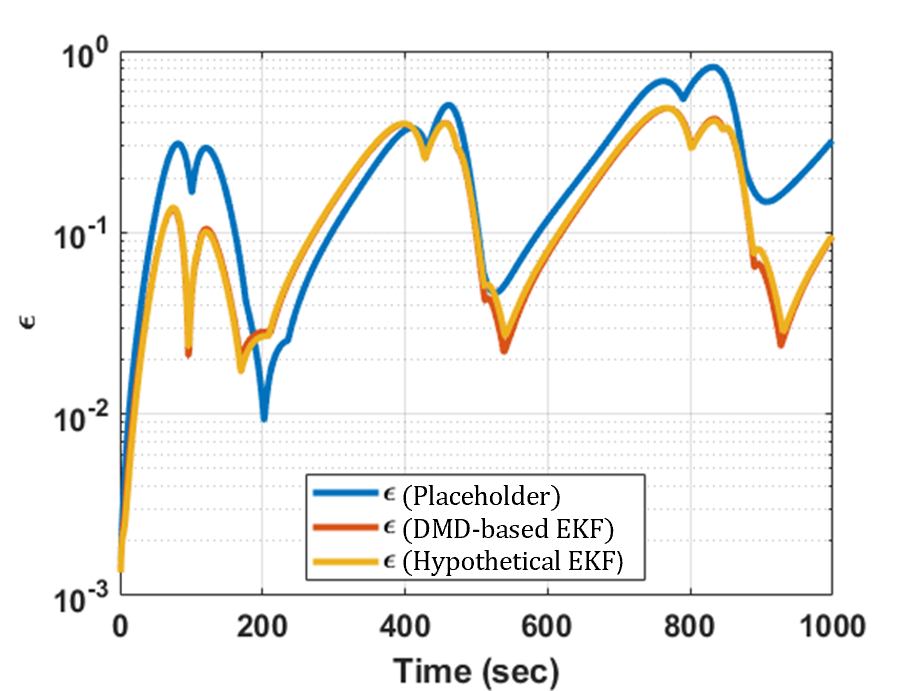}}{This figure shows the results of applying the proposed framework to the Van Der Pol oscillator with a true damping parameter of mu = 2 and a placeholder value of mu = 1.}
        \label{fig:vdp_pobs}
    }
    \subfloat[VDP Oscillator Results with Different $\mu$]{%
        \pdftooltip{\includegraphics[width=0.45\textwidth]{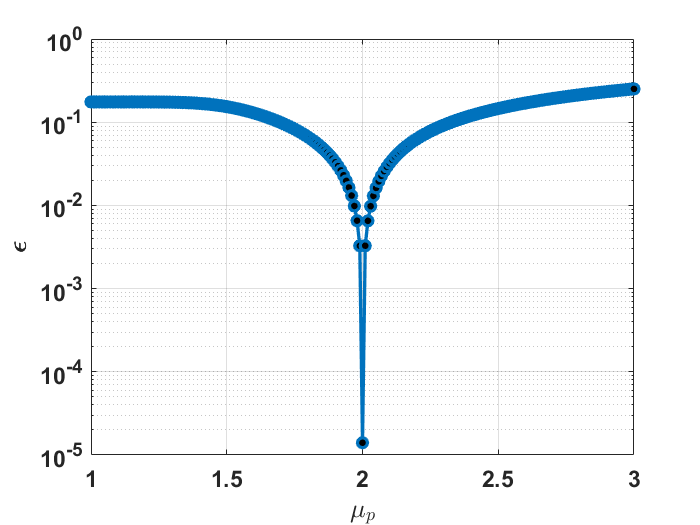}}{This figure shows the performance of the proposed framework for the Van Der Pol oscillator when sweeping over the placeholder damping parameter mu_p in the range [1,3].}
        \label{fig:vdp_pobs_mu}
    }
    \caption{VDP Oscillator Results}
    \label{fig:vdp_combined}
\end{figure}

This section addresses the issue of \textit{Model Assumptions}. In the previous scenarios, we considered a baseline model free of perturbations and a true system subject to perturbations and sensor noise. The framework then combined these two sources to produce reliable estimates. These scenarios included the pendulum with an unknown damping force, non-Keplerian motion with drag or J2 perturbations, and generalized perturbation models such as SGP4 and SDP4. 
In this section, we explore a different application where the placeholder model has entirely different parameters. For instance, consider the Van Der Pol Oscillator with a true damping parameter of $\mu = 2$, while the placeholder value used is $\mu = 1$. Noise-free position-only measurements are considered. In this oscillator, both parameters produce limit cycle behavior; however, the trajectories are vastly different. In this case, the framework will fail to produce reliable estimates. 
Fig.~\eqref{fig:vdp_pobs} illustrates this issue, showing that both the DMD-based EKF and the hypothetical EKF perform poorly. Although the system is periodic and Hankel-DMD accurately learns the measurements, the placeholder model is significantly different from the true system. Consequently, the proposed correction method is insufficient for this scenario.
To further justify, Fig.~\eqref{fig:vdp_pobs_mu} presents the performance of the framework when sweeping over the parameter $\mu_p \in [1,3]$ at 0.1 intervals; $\mu_p$ represents the damping parameter value for the placeholder system. It is seen that as the parameter value is in the close vicinity of the true parameter value, the error in the estimates is very small. However, as this guess is further off, it is seen that the error is orders of magnitude larger, and the estimates are no longer useful. This scenario highlights a particularly important case where the reference model dynamics are crucial for the success of this framework.
The Van Der Pol Oscillator example demonstrates the importance of having a reasonably accurate placeholder model. When the placeholder model's parameters deviate significantly from the true system, the fusion process in the EKF cannot compensate for the discrepancies, leading to poor state estimation. This limitation emphasizes the need for careful selection and validation of the reference dynamics model to ensure it provides a sufficiently accurate approximation of the true system's behavior. In practical applications, this might involve iterative refinement of the placeholder model parameters or the use of more sophisticated modeling techniques to better capture the system's dynamics.

\subsection{Extended Applications - Cislunar case}
\label{subsec:cislunar_case}

In this section, we extend the application of the proposed framework to the cislunar environment, which presents unique challenges due to the complex gravitational interactions between the Earth, Moon, and other celestial bodies. Accurate state estimation in this region is crucial for mission planning, navigation, and collision avoidance for spacecraft operating in or transiting through cislunar space. We demonstrate the effectiveness of our framework in this context by applying it to an example scenario. The results highlight the robustness and adaptability of the framework in handling the intricate dynamics of the cislunar environment, providing reliable state estimates even with partial and noisy observations.

\subsubsection{CR3BP - using noisy position measurements}
\label{subsubsec:cr3bp_pos}

\begin{table}[htbp]
    \fontsize{10}{10}\selectfont
    \caption{L1 - Halo orbit initial conditions in the rotating coordinate frame (nondimensionalized)}
    \label{tab:L1_init}
    \centering 
    \begin{tabular}{cccccc} 
        \toprule 
        $\mbf{x}$ & $\mbf{y}$ & $\mbf{z}$ & $\dot{\mbf{x}}$ & $\dot{\mbf{y}}$ & $\dot{\mbf{z}}$ \\
        \midrule 
        8.7592e-1 & -1.5903e-26 & 1.9175e-1 & -2.9302e-14 & 2.3080e-1 & 7.36497e-14 \\
        \bottomrule
    \end{tabular}
\end{table}

\begin{figure}
    \centering
    \subfloat[L1 HALO with Noise]{%
        \pdftooltip{\includegraphics[width=0.45\textwidth]{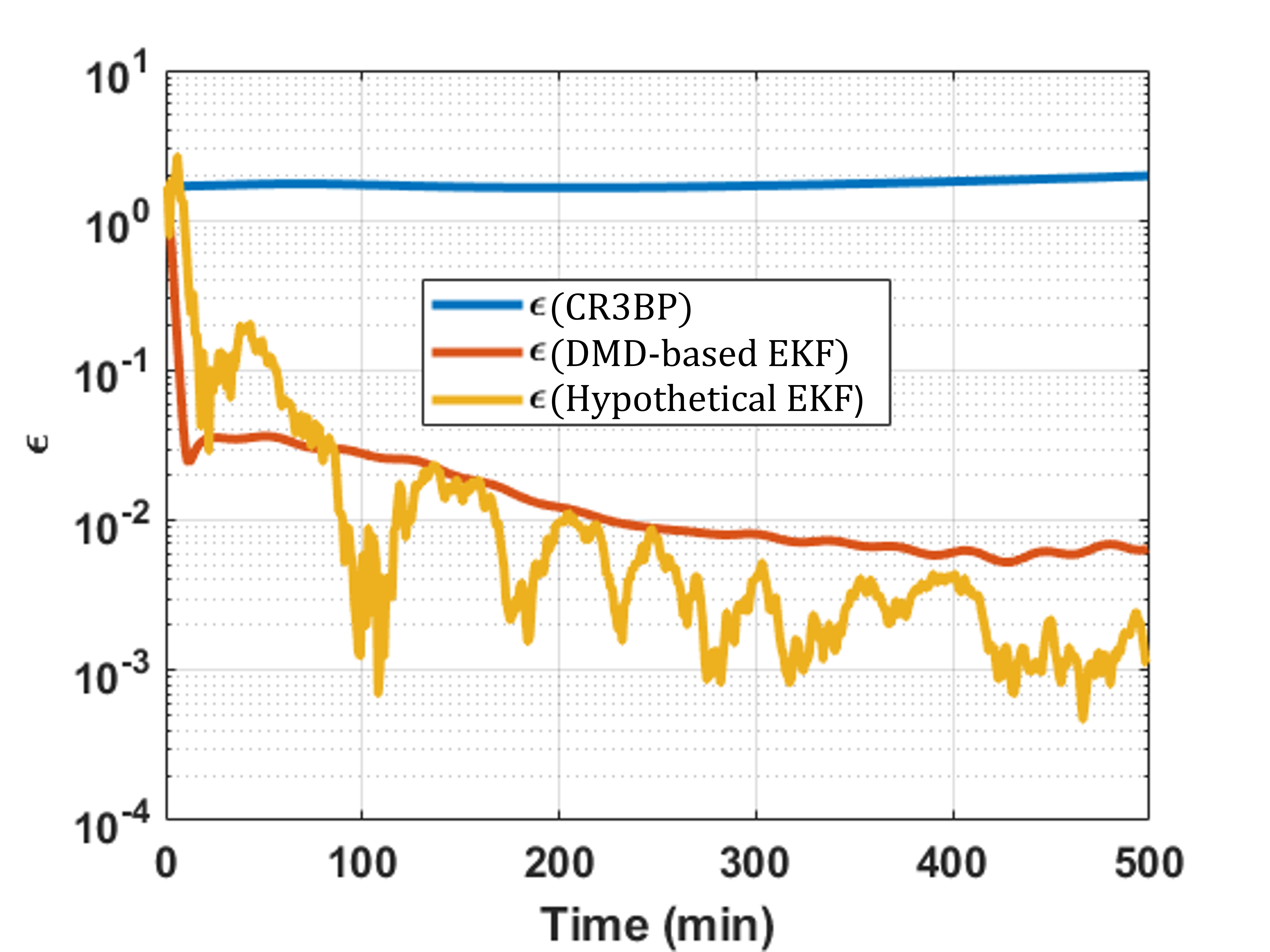}}{This figure shows the results of applying the proposed framework to an L1 HALO orbit with noisy position-only observations.}
        \label{fig:l1halo_pobs_pos_0}
    } \\
    \subfloat[L1 HALO with Noise]{%
        \pdftooltip{\includegraphics[width=0.45\textwidth]{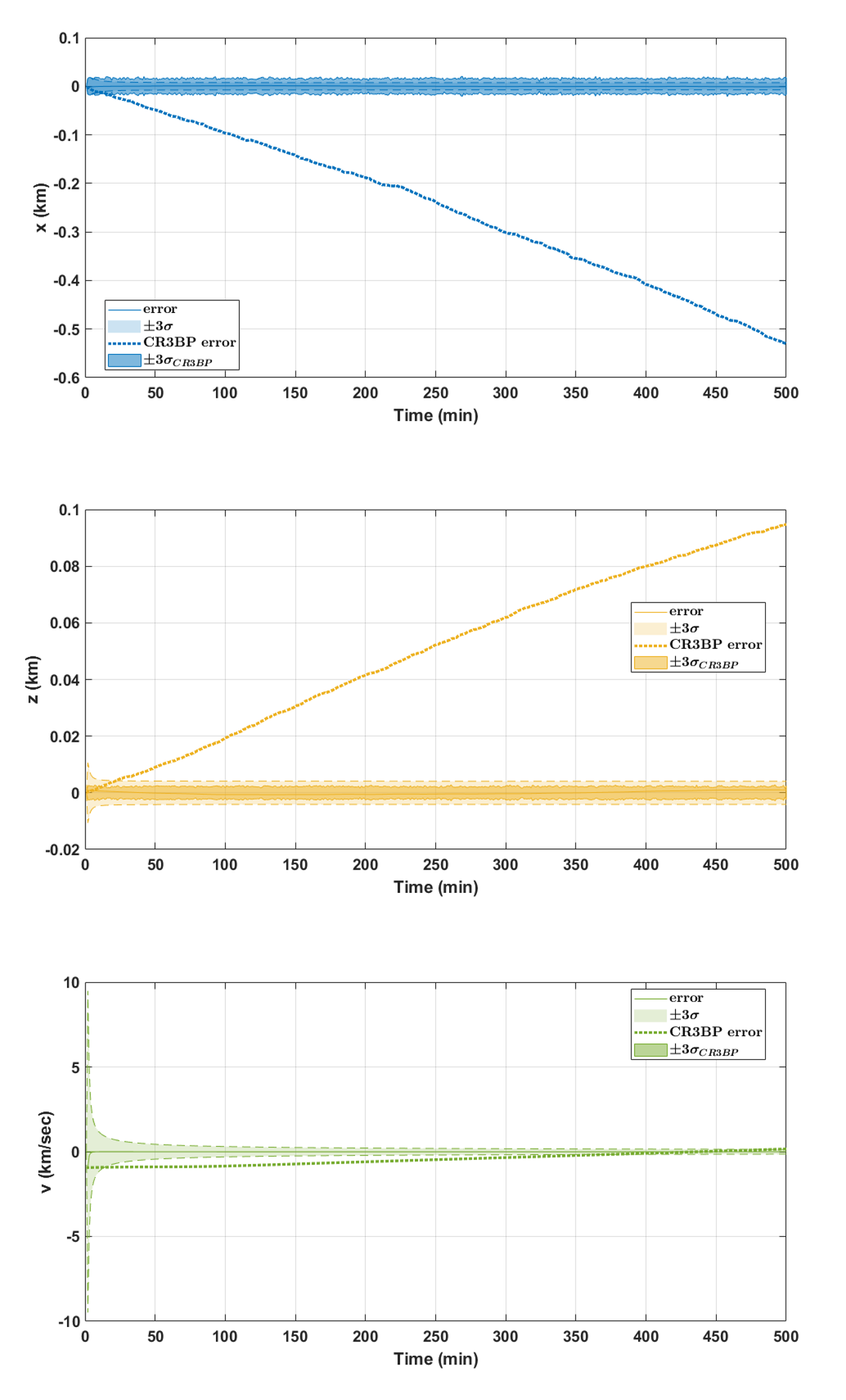}}{This figure shows the 3-sigma bounds for the x, z, and v state vectors of the L1 HALO orbit with noisy position-only observations.}
        \label{fig:l1halo_pobs_pos_1}
    }
    \subfloat[L1 HALO with Noise]{%
        \pdftooltip{\includegraphics[width=0.45\textwidth]{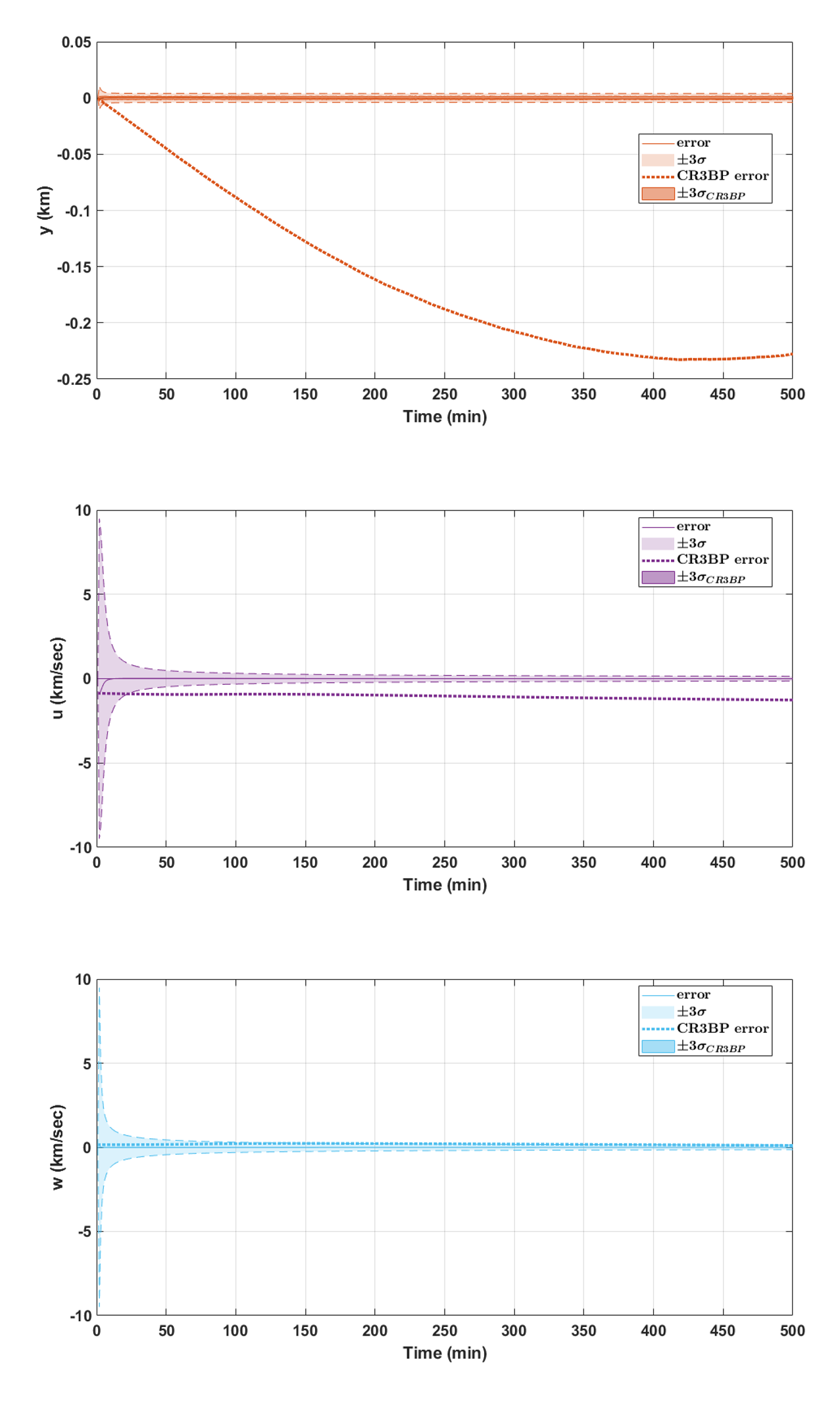}}{This figure shows the 3-sigma bounds for the y, u, and w state vectors of the L1 HALO orbit with noisy position-only observations.}
        \label{fig:l1halo_pobs_pos_2}
    }
    \caption[L1 HALO Results with Noise]{L1 HALO Results with Noise. This figure compares the performance of the proposed framework under noisy position-only observations for an L1 HALO orbit.}
    \label{fig:l1halo_results_with_noise}
\end{figure}

This section considers the application of the proposed framework to noisy position-only observations of an object in an L1-Halo orbit within the Cislunar environment. The initial conditions for the object are presented in Table.~\ref{tab:L1_init}. The placeholder model employed is the \ac{cr3bp} model, which does not account for any perturbations or noise. The measurements, however, are derived from the same \ac{cr3bp} dynamics but are overlaid with $1\%$ Gaussian noise in each state variable to simulate realistic sensor inaccuracies.
The framework is applied similarly to that described in previous sections. The initial conditions for the velocity components are determined using a finite difference approach. The covariance matrix and process noise parameters are initialized through a series of trial and error sweeps to ensure optimal performance. The sensor noise covariance matrix is directly derived from the noise characteristics of the measurements.
Figure~\eqref{fig:l1halo_results_with_noise} presents the results from this scenario. It is observed that the proposed framework effectively filters out the noise and performs almost as well as the hypothetical scenario where true measurements are available. Figures~\eqref{fig:l1halo_pobs_pos_1} and~\eqref{fig:l1halo_pobs_pos_2} illustrate the state vector estimates within the 3-$\sigma$ bounds, demonstrating that the estimates lie within this confidence region. This indicates the robustness of the framework in handling noisy observations.
In contrast, using the placeholder model alone without the corrections provided by the surrogate model and the EKF fusion process results in diverging behavior in the position states. Although the magnitude of this divergence is relatively small, on the order of meters, it can be insignificant when considering the scale of the \ac{cr3bp} problem, which involves distances on the order of hundreds of thousands of kilometers. 
In this scenario, the framework demonstrates its capability to effectively filter out noise and provide reliable state estimates, even with partial and noisy observations. 

\section{Assessing Consistency in Hankel-DMD-Based Surrogate Modeling}
\label{sec:consistency_analysis}

Since Hankel-DMD serves as the primary driver algorithm for extrapolating measurements, this section analyzes its consistency when fitting measurements $\mbf{z}(k)$. As it performs an autoregressive fit, we have,
\begin{align}
    \mbf{z}(k) = \alpha_1 \mbf{z}(k-1) + \alpha_2 \mbf{z}(k-2) + \cdots + \alpha_l \mbf{z}(k-l) + \mbf{v}(k),
\end{align}
where $\mbf{v} \sim \mathcal{N}(\mbf{0}, \mbf{R}(k))$. In matrix form, this can be written as,
\begin{align}
    \begin{bmatrix}
        \mbf{z}(k) \\
        \mbf{z}(k-1) \\
        \vdots \\
        \mbf{z}(k-l+1)
    \end{bmatrix}
    =
    \begin{bmatrix}
        \alpha_1 & \alpha_2 & \cdots & \alpha_l \\
        1 & 0 & \cdots & 0 \\
        \vdots & \vdots & \ddots & \vdots \\
        0 & \cdots & 1 & 0
    \end{bmatrix}
    \begin{bmatrix}
        \mbf{z}(k-1) \\
        \mbf{z}(k-2) \\
        \vdots \\
        \mbf{z}(k-l)
    \end{bmatrix}
    +
    \begin{bmatrix}
        \mbf{v}(k) \\
        0 \\
        \vdots \\
        0
    \end{bmatrix}.
\end{align}

Considering $\mbf{X}_{k+1} \approx {\mbf{A}\mbf{X}_k}$ and the Hankel formulation from Ref.~\cite{narayanan2024predictive}, we have,
\begin{align}
    \begin{bmatrix}
        \mbf{z}(k) & \mbf{z}(k-1) & \cdots & \mbf{z}(k-m+1)\\
        \mbf{z}(k-1) & \mbf{z}(k-2) & \cdots & \vdots\\
        \vdots & \vdots & \ddots & \vdots\\
        \mbf{z}(k-l+1) & \mbf{z}(k-l) & \cdots & \mbf{z}(k-m-l+2)
    \end{bmatrix}
    = \nonumber \\
    \mbf{A}
    \begin{bmatrix}
        \mbf{z}(k-1) & \mbf{z}(k-2) & \cdots & \mbf{z}(k-m)\\
        \mbf{z}(k-2) & \mbf{z}(k-3) & \cdots & \vdots\\
        \vdots & \vdots & \ddots & \vdots\\
        \mbf{z}(k-l) & \mbf{z}(k-l-1) & \cdots & \mbf{z}(k-m-l+1)
    \end{bmatrix} \nonumber \\
    + 
    \begin{bmatrix}
        \mbf{v}(k-1) \\
        \mbf{v}(k-2) \\
        \vdots \\
        \mbf{v}(k-m)
    \end{bmatrix},
\end{align}
thus giving the equation,
\begin{align}
    \mbf{z}_{k+1} \approx \mbf{A}\mbf{z}_k + \mbf{V}_k.
\end{align}

Using Hankel-DMD, the best-fit linear operator $\mbf{A}$ is obtained by minimizing the cost function $J$,
\begin{align}
    \min_{\mbf{A}} J &= \frac{1}{2} (\mbf{z}_{k+1} - \mbf{A} \mbf{z}_k - \mbf{V}_k)^\top (\mbf{z}_{k+1} - \mbf{A} \mbf{z}_k - \mbf{V}_k).
\end{align}
Taking the derivative of $J$ with respect to $\mbf{A}$ and setting it to zero, we get,
\begin{align}
    \frac{\partial J}{\partial \mbf{A}} &= 0 \\
    \implies \mbf{z}_k (\mbf{z}_{k+1} - \mbf{A} \mbf{z}_k - \mbf{V}_k)^\top &= 0 \\
    \implies \mbf{z}_k \mbf{z}_{k+1}^\top - \mbf{A} \mbf{z}_k \mbf{z}_k^\top &= 0 \\
    \implies \mbf{A} &= \mbf{z}_{k+1} \mbf{z}_k^\top (\mbf{z}_k \mbf{z}_k^\top)^{-1}.
\end{align}

Taking the expectation of $\mbf{A}$, we have,
\begin{align}
    \mathbb{E}[\mbf{A}] &= \mathbb{E}\left[\mbf{z}_{k+1} \mbf{z}_k^\top (\mbf{z}_k \mbf{z}_k^\top)^{-1}\right].
\end{align}

Since $\mbf{z}_{k+1} = \mbf{A} \mbf{z}_k + \mbf{V}_k$, we can substitute this into the expectation,
\begin{align}
    \mathbb{E}[\mbf{A}] &= \mathbb{E}\left[(\mbf{A} \mbf{z}_k + \mbf{V}_k) \mbf{z}_k^\top (\mbf{z}_k \mbf{z}_k^\top)^{-1}\right].
\end{align}

Expanding the expectation, we get,
\begin{align}
    \mathbb{E}[\mbf{A}] &= \mathbb{E}\left[\mbf{A} \mbf{z}_k \mbf{z}_k^\top (\mbf{z}_k \mbf{z}_k^\top)^{-1}\right] + \mathbb{E}\left[\mbf{V}_k \mbf{z}_k^\top (\mbf{z}_k \mbf{z}_k^\top)^{-1}\right].
\end{align}

Since $\mathbb{E}[\mbf{V}_k] = 0$, the second term vanishes,
\begin{align}
    \mathbb{E}[\mbf{A}] &= \mathbb{E}\left[\mbf{A} \mbf{z}_k \mbf{z}_k^\top (\mbf{z}_k \mbf{z}_k^\top)^{-1}\right].
\end{align}

Simplifying, we get,
\begin{align}
    \mathbb{E}[\mbf{A}] &= \mbf{A} \mathbb{E}\left[\mbf{z}_k \mbf{z}_k^\top (\mbf{z}_k \mbf{z}_k^\top)^{-1}\right].
\end{align}

Since $\mbf{z}_k \mbf{z}_k^\top (\mbf{z}_k \mbf{z}_k^\top)^{-1} = \mbf{I}$, we have,
\begin{align}
    \mathbb{E}[\mbf{A}] &= \mbf{A} \mathbb{E}[\mbf{I}] = \mbf{A}.
\end{align}

Therefore, $\mbf{A}$ is unbiased. 
To analyze the estimation error of the extrapolated measurements, we define 
\begin{equation}
    \mbf{e} = \mbf{z}_{k+1} - \mbf{A} \mbf{z}_k.
\end{equation}

The mean of the estimation error $\mbf{e}$ is given by,
\begin{align}
    \mathbb{E}[\mbf{e}] &= \mathbb{E}[\mbf{z}_{k+1} - \mbf{A} \mbf{z}_k] \\
    &= \mathbb{E}[\mbf{z}_{k+1}] - \mathbb{E}[\mbf{A} \mbf{z}_k].
\end{align}

Since $\mbf{z}_{k+1} = \mbf{A} \mbf{z}_k + \mbf{V}_k$, we have,
\begin{align}
    \mathbb{E}[\mbf{e}] &= \mathbb{E}[\mbf{A} \mbf{z}_k + \mbf{V}_k] - \mathbb{E}[\mbf{A} \mbf{z}_k] \\
    &= \mathbb{E}[\mbf{A} \mbf{z}_k] + \mathbb{E}[\mbf{V}_k] - \mathbb{E}[\mbf{A} \mbf{z}_k] \\
    &= \mathbb{E}[\mbf{V}_k].
\end{align}

Since $\mathbb{E}[\mbf{V}_k] = \mbf{0}$, we have,
\begin{equation}
    \mathbb{E}[\mbf{e}] = \mbf{0}.
\end{equation}

The variance of the estimation error $\mbf{e}$ is given by,
\begin{align}
    \text{Var}(\mbf{e}) &= \mathbb{E}[(\mbf{e} - \mathbb{E}[\mbf{e}])(\mbf{e} - \mathbb{E}[\mbf{e}])^\top] \\
    &= \mathbb{E}[\mbf{e} \mbf{e}^\top] \\
    &= \mathbb{E}[(\mbf{z}_{k+1} - \mbf{A} \mbf{z}_k)(\mbf{z}_{k+1} - \mbf{A} \mbf{z}_k)^\top].
\end{align}

Substituting $\mbf{z}_{k+1} = \mbf{A} \mbf{z}_k + \mbf{V}_k$, we get,
\begin{align}
    \text{Var}(\mbf{e}) &= \mathbb{E}[(\mbf{A} \mbf{z}_k + \mbf{V}_k - \mbf{A} \mbf{z}_k)(\mbf{A} \mbf{z}_k + \mbf{V}_k - \mbf{A} \mbf{z}_k)^\top] \\
    &= \mathbb{E}[\mbf{V}_k \mbf{V}_k^\top] + \mathbb{E}[\mbf{A} \mbf{z}_k \mbf{z}_k^\top \mbf{A}^\top] - \mathbb{E}[\mbf{A} \mbf{z}_k \mbf{V}_k^\top] - \mathbb{E}[\mbf{V}_k \mbf{z}_k^\top \mbf{A}^\top].
\end{align}

Since $\mbf{V}_k \sim \mathcal{N}(\mbf{0}, \mbf{R}(k))$ and $\mathbb{E}[\mbf{V}_k \mbf{z}_k^\top] = 0$, we have,
\begin{equation}
    \text{Var}(\mbf{e}) = \mbf{R}(k) + \mbf{A} \mathbb{E}[\mbf{z}_k \mbf{z}_k^\top] \mbf{A}^\top.
    \label{eq:var_estimate}
\end{equation}

The EKF framework relies on a good initial estimate for the sensor noise covariance $\mbf{R}$. However, Eq.~\eqref{eq:var_estimate} shows that $\text{Var}(\mbf{e}) > \mbf{R}$, and over time, this value will grow due to the additional term $\mbf{A} \mathbb{E}[\mbf{z}_k \mbf{z}_k^\top] \mbf{A}^\top$. Further, in real applications, achieving a fit close to machine precision is not feasible. This is evident in the error magnitude charts, where the relative error typically ranges from $10^{-3}$ to $10^{-6}$. Consequently, the covariance grows over time due to the inability to extrapolate accurately.
To address this issue, it is essential to update the surrogate model with new measurements periodically. This can be achieved by incorporating a mechanism to retrain the Hankel-DMD model whenever new measurement data becomes available. By doing so, the surrogate model can adapt to changes in the system dynamics and maintain its accuracy over time.
Moreover, the process noise covariance $\mbf{Q}$ should be carefully tuned to account for the uncertainties in the system dynamics. A well-chosen $\mbf{Q}$ can help balance model predictions with actual measurements, thereby improving the overall performance of the EKF framework.
While this work focuses on one-shot learning, future research will explore scenarios where new measurements are available, how they are handled, and their implications when using an iterative training approach. The following section discusses the results presented in this work.

\section{Summary}
\label{sec:summary_5}
This work introduced a framework for forecasting with partial observations using filtering. To improve state estimation, the surrogate models and reference dynamics were fused via an EKF. Applications to systems like a pendulum, ISS, and cislunar orbits demonstrated its effectiveness, even with noisy or limited data. Challenges such as surrogate model accuracy, noise sensitivity, and extrapolation limits were highlighted through case studies like Molniya-3-50 and Van der Pol oscillator. The framework's robustness in complex environments, such as cislunar space, was validated. Consistency analysis emphasized the need for periodic surrogate model updates to maintain accuracy.
The consistency analysis in this work is a preliminary step toward validating the framework's robustness. Future work will focus on formal theoretical guarantees, handling multiple measurements, and addressing potential biases like "double dipping" when retraining surrogate models. These efforts aim to enhance the framework's reliability and applicability.
In summary, the proposed framework effectively integrates surrogate models and reference dynamics to improve state estimation with partial observations. It demonstrates robustness across various scenarios, including noisy and complex environments like cislunar space, making it a versatile tool for diverse applications.

\bibliographystyle{AAS_publication}   
\bibliography{bibfile}   

\end{document}